\documentclass[superscriptaddress,twocolumn,english,floatfix,aps,prd,amsmath,amssymb,longbibliography]{revtex4-2}
\usepackage[T1]{fontenc}
\usepackage[latin9]{inputenc}
\usepackage{color}
\usepackage{times}
\usepackage{babel}
\usepackage{epsfig}
\usepackage{subfigure}
\usepackage{gensymb}
\usepackage{textcomp}
\usepackage{amsmath}
\usepackage{amssymb}
\usepackage{amsbsy}
\usepackage{graphicx}
\usepackage{float}
\usepackage{esint}
\usepackage[colorlinks=true]{hyperref}
\hypersetup{citecolor=blue,linkcolor=blue,urlcolor=blue}

\begin{document}
	

\title{Introducing corrugated surfaces in electromagnetism problems via perturbative approach}

\author{Alexandre P. da Costa}
\email{alexandre.pereira.costa@icen.ufpa.br}
\affiliation{Faculdade de F\'{i}sica, Universidade Federal do Par\'{a}, 66075-110, Bel\'{e}m, Par\'{a}, Brazil}

\author{Lucas Queiroz}
\email{lucas.silva@icen.ufpa.br}
\affiliation{Faculdade de F\'{i}sica, Universidade Federal do Par\'{a}, 66075-110, Bel\'{e}m, Par\'{a}, Brazil}

\author{Edson C. M. Nogueira}
\email{edson.moraes.nogueira@icen.ufpa.br}
\affiliation{Faculdade de F\'{i}sica, Universidade Federal do Par\'{a}, 66075-110, Bel\'{e}m, Par\'{a}, Brazil}

\author{Danilo T. Alves}
\email{danilo@ufpa.br}
\affiliation{Faculdade de F\'{i}sica, Universidade Federal do Par\'{a}, 66075-110, Bel\'{e}m, Par\'{a}, Brazil}
\affiliation{Centro de F\'{i}sica, Universidade do Minho, P-4710-057, Braga, Portugal}
\date{\today}
	
\begin{abstract}
	Problems involving boundary conditions on corrugated surfaces are relevant to understand nature,
	since, at some scale, surfaces manifest corrugations that have to be taken into account.
	In introductory level electromagnetism courses, a very common and fundamental exercise is
	to solve Poisson's equation for a point charge in the presence of an infinity
	perfectly conducting planar surface, which is usually done by image method. 
	Clinton, Esrick and Sacks [Phys. Rev. B 31, 7540 (1985)] added corrugation to this surface, and
	solved the problem by a perturbative analytical calculation of the corresponding Green's function.
	In the present paper, we make a detailed pedagogical review of this calculation, 
	aiming to popularize their results.
	We also present an original contribution, extending this perturbative approach
	to solve the Laplace's equation for the electrostatic potential for a corrugated neutral conducting cylinder in the presence of a uniform electric field (without corrugation, this is another very common model considered as an exercise in electromagnetism courses).
	All these calculations can be used as pedagogical examples of the application of the present perturbative approach in electromagnetism courses.
\end{abstract}

	\maketitle

\section{Introduction}
\label{Intro-geral}

In many physical situations of interaction between bodies, in a certain scale of length, the corrugation of their surfaces have to be taken into account,
for instance in the calculation of the image potential of a charge in the vicinity of a surface \cite{Clinton-PRBI-1985, Rahman-PRB-1980}. 
The image potential, which is the potential that arises from the induced charge distribution, is important for several effects, such as the formation of image potential states, which are quantum states of electrons localized at the surface of materials that exhibit negative electron affinity \cite{Rahman-PRB-1980}. 
These electrons are bound close to the surface due to the image potential field, which attracts them to the surface, but up to a certain point after which a repulsive force does not allow the electron to penetrate the material \cite{Echenique-SufSci-1991}. 
The behavior of these electrons can reveal important physical and chemical surface information \cite{Cole-RMP-1974,Hong-PRB-2001}. 
The image potential presents modifications when the surface has deformations \cite{Clinton-PRBI-1985, Hong-PRB-2001} and, since the fabrication of perfectly plane surfaces is almost impossible, the corrugation effects in the image potential may have to be taken into account 
\cite{Rahman-PRB-1980}.

Until 1980, the majority of the investigations on the image potential effects were made 
considering that the interfaces between two dielectric media were planar \cite{Rahman-PRB-1980}. 
In 1985, Clinton, Esrick and Sacks calculated the image potential for a point charge in 
vacuum in the presence of a nonplanar metallic surface \cite{Clinton-PRBI-1985}.  
They showed that ions and electrons are always attracted to the elevated part of the surface \cite{Clinton-PRBI-1985}.
Later, also in 1985, these authors, considering corrugation in an infinite conducting plane, 
solved, by perturbative analytical calculations, the Poisson's equation for a point charge in the 
presence of  such a corrugated surface \cite{Clinton-PRBII-1985}.
This approach has been applied recently,
for instance,  in Ref. \cite{Alves-PRB-2019}, where
it was considered a planar two-dimensional system (for instance,  graphene or transition metal 
dichalcogenide monolayers) in the presence of a nonplanar dielectric medium and,
using an extension of the perturbative method found in Ref. \cite{Clinton-PRBII-1985},
it was obtained that the effective potential of the electron-electron interaction in a 2D system is dependent 
not only on the distance to the source charge, but also on the position of the charge itself.
As another example, in Ref. \cite{Nogueira-PRA-2021} the authors combined the perturbative analytical solution in Ref. \cite{Clinton-PRBII-1985} with the description done by Eberlein and Zietal \cite{Eberlein-PRA-2007} for the van der Waals interaction between a polarizable particle and an ideal conducting surface, proposing 
a new analytical approach to investigate the van der Waals interaction between an anisotropic particle and a corrugated surface.
In this way, in Ref. \cite{Nogueira-PRA-2021} it was predicted new nontrivial behaviors of the lateral van der Waals force, so that the particle can be attracted not only toward the corrugation peaks (as found so far in the literature), but also to the nearest valley, or to an intermediate point between a peak and a valley \cite{Nogueira-PRA-2021}.
Calculations based on Ref. \cite{Clinton-PRBII-1985}
have been used to extend the results shown in Ref. \cite{Nogueira-PRA-2021} 
to dielectrics \cite{Queiroz-PRA-2021}.
Moreover, in Ref. \cite{Nogueira-Arxiv-2021}, 
it was predicted, taking as basis Ref. \cite{Clinton-PRBII-1985}, 
the possibility of sign inversions in the lateral van
der Waals force acting on an anisotropic particle interacting with a 
conducting plane containing a single  protuberance.
 
In introductory level electromagnetism courses, a very common and fundamental exercise is to solve Poisson's equation for a point charge in the presence of an infinity perfectly conducting planar surface \cite{Griffiths, Jackson, Machado, Milford, Purcell, Nayfeh-Electrodynamics-1985}, which is usually done by image method.
Clinton, Esrick and Sacks \cite{Clinton-PRBII-1985} introduced corrugation to this surface, and solved the problem via a perturbative analytical calculation.
Thus, in the present paper, we make a detailed pedagogical review of their calculations, aiming to popularize their results and to use them as a pedagogical example of the introduction of corrugated surfaces in electromagnetism problems via a perturbative approach.
In addition, we give a special attention to the cases of a
sinusoidal surface,
and show how the potential, electric field, and induced surface charge density are affected by the presence of a corrugation on the surface.
We also present an original contribution, extending this perturbative approach
to solve the Laplace's equation for the electrostatic potential for a corrugated neutral conducting cylinder in the presence of a uniform electric field (without corrugation, this is another very common model considered as an exercise in electromagnetism courses \cite{Griffiths, Jackson, Machado, Milford, Purcell, Nayfeh-Electrodynamics-1985}).
This calculation can be used as another, and simpler, pedagogical example of the application of a perturbative approach to discuss the introduction of corrugated surfaces in electromagnetism problems.

This paper is organized as follows. 
In Sec. \ref{review-point-plane}, we make a brif review of the problem of a point charge in the presence of an infinity perfectly conducting planar surface. 
In Sec. \ref{review-clinton}, we make a detailed review of the Clinton, Esrick and Sacks perturbative analytical calculations to solve Poisson's equation in the presence of a nonplanar conducting surface. 
In Sec. \ref{translated-surface}, we perform, for pedagogical purposes, a consistency check of the formulas obtained in the previous section, by applying them to the case of a translated plane, and discuss some conditions under which the approximate solutions become closer to the exact one.
In Sec. \ref{sinusoidal-surface}, we perform an application of the Clinton, Esrick and Sacks calculations for a sinusoidal surface.
In Sec. \ref{normal-cylinder}, we review the solution for the Laplace's equation for an ideal conducting cylinder placed in a uniform electrostatic field.
In Sec. \ref{corrugated-cylinder}, we extend this perturbative approach
to solve the Laplace's equation for the electrostatic potential of an infinite conducting corrugated cylinder placed in a uniform electrostatic field.
In Sec. \ref{radius sigthly enhanced}, we perform a consistency check of the formulas obtained in the previous section, by applying them to the case of a cylinder with a slight enhancement in its radius.
In Sec. \ref{application Cylinder corrgated}, we apply our perturbative calculations to a sinusoidal corrugated cylinder, and analyze how the potential, electric field, and induced surface charge density are affected by the corrugation.
In Sec. \ref{final-remarks}, we present our final remarks.

\section{A point charge in the presence of an infinity planar conducting surface}
\label{review-point-plane}

\begin{figure}
\centering
\epsfig{file=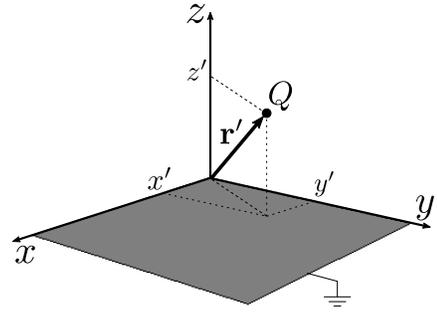,  width=0.65 \linewidth}
\caption{{
Illustration of a charge $Q$, located at ${\bf r}^{\prime}={\bf r}_{||}^{\prime}+z^\prime\hat{{\bf z}}$ (with $z^\prime>0$), 
in the presence of a grounded conducting plane.
}}
\label{fig:carga-superficie-plana}
\end{figure}

In this section, we make a brief review of the problem of finding the potential $\Phi$ for a point charge at a distance $z^{\prime}$ from an infinite grounded conducting plane located at $z=0$ (see Fig. \ref{fig:carga-superficie-plana}).
Let us consider a point charge $Q$ located at the position 
${\bf r}^{\prime}={\bf r}_{||}^{\prime}+z^\prime\hat{{\bf z}}$
(with $z^\prime>0$ and ${\bf r}_{||}^\prime=x^\prime\hat{{\bf x}}+y^\prime\hat{{\bf y}}$). 
The problem of finding a solution to $\Phi$, for this case, is a very common exercise in introductory electromagnetism courses (see, for instance, Refs. \cite{Griffiths, Jackson, Machado, Milford, Purcell, Nayfeh-Electrodynamics-1985}).
By writing the potential as
\begin{equation}
	\label{phi-to-green}
	\Phi\left({\bf r}, {\bf r}^{\prime}\right)=QG\left({\bf r},{\bf r}^{\prime}\right),
\end{equation}
where $G\left({\bf r},{\bf r}^{\prime}\right)$ is the Green's function of the Laplacian operator, one has that the problem consists in finding (in the region $z \geq 0$) the solution of Poisson's equation
\begin{equation}
	\boldsymbol{\nabla}^2{G\left({\bf r},{\bf r}^{\prime}\right)}=-4\pi \delta\left({\bf r}-{\bf r}^{\prime}\right),
	\label{poisson}
\end{equation}
under the boundary conditions
\begin{align}
	\left.G\left(\textbf{r},\textbf{r}^{\prime}\right)\right|_{z=0}&=0,
	\label{bc-plane}\\
	\left.G\left(\textbf{r},\textbf{r}^{\prime}\right)\right|_{|\textbf{r}|\gg |\textbf{r}^\prime|}&\to 0.
	\label{bc-plane-infinity}
\end{align}

\begin{figure}
	\centering
	\epsfig{file=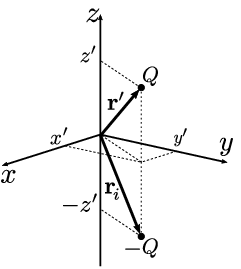,  width=0.5 \linewidth}
	\caption{{
			Illustration of a system formed by a charge $Q$, located at ${\bf r}^{\prime}={\bf r}_{||}^\prime+z^\prime\hat{{\bf z}}$
			(with $z^\prime>0$), and $-Q$ (image charge) located at 
			$\textbf{r}_i={\bf r}_{||}^{\prime}-z^\prime\hat{{\bf z}}$.
	}}
	\label{fig:carga-Q-Q-imagem}
\end{figure}

The solution of the problem given in Eqs. \eqref{poisson}-\eqref{bc-plane-infinity} 
is usually obtained by the image method \cite{Griffiths, Jackson, Machado, Milford, Purcell, Nayfeh-Electrodynamics-1985}.
This means that it coincides, for $z\geq 0$, with the solution of another 
problem, defined by the configuration formed by two point charges: $Q$ located at $\textbf{r}=\textbf{r}^{\prime}$, and $-Q$ located at $\textbf{r}=\textbf{r}_i$
(see Fig. \ref{fig:carga-Q-Q-imagem}), where $\textbf{r}_i={\bf r}_{||}^{\prime}-z^\prime\hat{{\bf z}}$ (the charge $-Q$
is named as the image charge of $Q$).
For this new configuration, the solution is simply given by the sum of the correspondent Green's function of each charge, and is written as
\begin{eqnarray}
	\nonumber
	G\left({\bf r}, {\bf r}^{\prime}\right)&=&\frac{{1}}{[|\textbf{r}_{\parallel}-\textbf{r}_{\parallel}^{\prime}|^{2}+\left(z-z^{\prime}\right)^{2}]^{\frac{1}{2}}}
	\\
	&-&\frac{{1}}{[|\textbf{r}_{\parallel}-\textbf{r}_{\parallel}^{\prime}|^{2}+\left(z+z^{\prime}\right)^{2}]^{\frac{1}{2}}}.
	\label{phi0}
\end{eqnarray}
This solution, for $z \geq 0$, is also the solution for the original problem given
by Eqs. \eqref{poisson}-\eqref{bc-plane-infinity} and illustrated in Fig. \ref{fig:carga-superficie-plana}.

\section{Review of Clinton, Esrick and Sacks calculations}
\label{review-clinton}

\begin{figure}
	\centering
	\epsfig{file=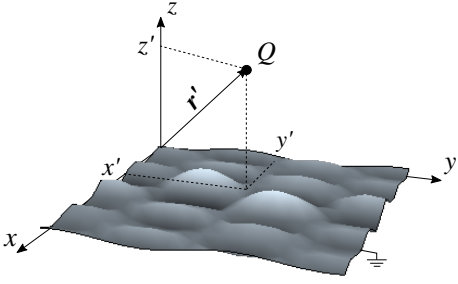,  width=0.7 \linewidth}
	\caption{{
			Illustration of a charge $Q$, located at ${\bf r}^{\prime}={\bf r}_{||}^\prime+z^\prime\hat{{\bf z}}$
			(with $z^\prime>0$), interacting with a general grounded conducting corrugated surface, whose corrugation profile is described by $z=h(\textbf{r}_\parallel)$.
	}}
	\label{fig:carga-superficie-geral}
\end{figure}

In this section, we make a brief pedagogical review of the Clinton, Esrick and Sacks perturbative analytical calculations and results found in Ref. \cite{Clinton-PRBII-1985}.
These authors computed, perturbatively, the potential $\Phi$ associated to a problem similar to that illustrated in Fig. \ref{fig:carga-superficie-plana}, with the difference that they considered a corrugated surface, instead of a plane.
In other words, they considered a point charge $Q$ located at ${\bf r}^{\prime}$, in the presence of a grounded conducting corrugated surface given by $z=h({\bf r}_{||})$ (see Fig. \ref{fig:carga-superficie-geral}), where $h({\bf r}_{||})$ describes a suitable modification $[\text{max}|h({\bf r}_{||})|\ll z^\prime]$ of a grounded planar conducting surface at $z=0$.
Effectively, one has that the problem consists in finding [in the region $z\geq h({\bf r}_{||})$] the solution of Eq. \eqref{poisson}, under the condition that $G\left(\textbf{r},\textbf{r}^{\prime}\right)$ goes to zero at large distances [Eq. \eqref{bc-plane-infinity}], and the boundary condition
\begin{equation}
\left.G\left(\textbf{r},\textbf{r}^{\prime}\right)\right|_{z=h({\bf r}_{||})}=0.
\label{bc-corrugated}
\end{equation}

To build a perturbative description, these authors introduced
an arbitrary auxiliary parameter $\varepsilon$, with $0\leq \varepsilon \leq 1$. 
Thus, one parameterize: $h({\bf r}_{||})\to \varepsilon h({\bf r}_{||})$, so that
Eq. \eqref{bc-corrugated} is replaced by
\begin{equation}
	\left.G\left(\textbf{r},\textbf{r}^{\prime}\right)\right|_{z=\varepsilon h({\bf r}_{||})}=0,	
	\label{bc-corrugated-epsilon}
\end{equation}
and the Green's function is written as a perturbative expansion in terms of the parameter $\varepsilon$ \cite{Clinton-PRBII-1985},
\begin{eqnarray}
	\label{Green_Gen}
	G(\textbf{r},\textbf{r}^{\prime})=G^{(0)}(\textbf{r},\textbf{r}^{\prime})+\sum_{n=1}^{\infty}\varepsilon^{n}G^{(n)}(\textbf{r},\textbf{r}^{\prime}),
\end{eqnarray}   
where $G^{(0)}(\textbf{r},\textbf{r}^{\prime})$ is the unperturbed Green's function [Eq. \eqref{phi0}, now considered for $z\geq h(\textbf{r}_\parallel)$] corresponding to the configuration obtained by image method and illustrated in Fig. \ref{fig:carga-Q-Q-imagem}, and $G^{(n)}(\textbf{r},\textbf{r}^{\prime})$ are the perturbative corrections. 
Note that, when $\varepsilon=0$, one has the correspondent solution to the case of a planar surface at $z=0$, whereas, when $\varepsilon=1$, we recover the actual problem of a corrugated surface described by $z=h(\textbf{r}_\parallel)$.
Thus, when we find $G^{(0)}(\textbf{r},\textbf{r}^{\prime})$ and $G^{(n)}(\textbf{r},\textbf{r}^{\prime})$, one must set $\varepsilon=1$ to obtain the solution for the problem.

Substituting Eq. \eqref{Green_Gen} in Eq. \eqref{poisson}, one obtains
\begin{eqnarray}
	\sum_{n=0}^{\infty}\varepsilon^{n}
	\left[\boldsymbol{\nabla}^2 G^{(n)}(\textbf{r},\textbf{r}^{\prime})\right]
	=-4\pi \delta\left({\bf r}-{\bf r}^{\prime}\right),
	\label{poisson-n}
\end{eqnarray}   
which can be rewritten as
\begin{align}
	\boldsymbol{\nabla}^2 G^{(0)}(\textbf{r},\textbf{r}^{\prime})
	+4\pi \delta\left({\bf r}-{\bf r}^{\prime}\right)
	+\sum_{n=1}^{\infty}\varepsilon^{n}
	\left[\boldsymbol{\nabla}^2 G^{(n)}(\textbf{r},\textbf{r}^{\prime})\right]
	=0.
	\label{poisson-n-expandido}
\end{align}   
Since $\varepsilon$ is an arbitrary parameter, which can be chosen from $0$ to $1$, the solution of Eq. \eqref{poisson-n-expandido} requires that the coefficients multiplying each power of $\varepsilon$ vanish.
In this way, one obtains the differential equations to $G^{(0)}$ and each $G^{(n)}$, which are given by
\begin{eqnarray}
	\boldsymbol{\nabla}^2 G^{(0)}(\textbf{r},\textbf{r}^{\prime})
	&=&-4\pi \delta\left({\bf r}-{\bf r}^{\prime}\right),
	\label{poisson-ordem-0}
	\\
	\boldsymbol{\nabla}^2 G^{(n)}(\textbf{r},\textbf{r}^{\prime})&=&0 \quad (n\geq1).
	\label{poisson-ordem-n}
\end{eqnarray}   

Using, in Eq. \eqref{Green_Gen}, the boundary condition given by Eq. \eqref{bc-corrugated-epsilon}, one obtains
\begin{eqnarray}
	\label{conection}
	\sum_{i=0}^{\infty}\varepsilon^{i}G^{(i)}(\textbf{r},\textbf{r}^{\prime})\vert_{z=\varepsilon h(\textbf{r}_\parallel)}=0.
\end{eqnarray}
Expanding it in powers of $\varepsilon h(\textbf{r}_\parallel)$, one has \cite{Clinton-PRBII-1985}
\begin{eqnarray}
	\sum_{i,m=0}^{\infty}\varepsilon^{i+m}\left[h(\textbf{r}_\parallel)\right]^{m}\frac{1}{m!}\frac{\partial^{m}}{\partial z^{m}}G^{(i)}(\textbf{r},\textbf{r}^{\prime})\vert_{z=0}=0.
\end{eqnarray}
Manipulating the series to bring together the terms with a same order in $\varepsilon$, one obtains
\begin{widetext}
\begin{align}
\nonumber
&G^{(0)}(\textbf{r},\textbf{r}^{\prime})\vert_{z=0}+\varepsilon\left[h(\textbf{r}_\parallel)\frac{\partial}{\partial z}G^{(0)}(\textbf{r},\textbf{r}^{\prime})\vert_{z=0}+G^{(1)}(\textbf{r},\textbf{r}^{\prime})\vert_{z=0}\right] 
\\
\nonumber 
&+\varepsilon^{2}\left\{\left[h(\textbf{r}_\parallel)\right]^{2}\frac{1}{2}\frac{\partial^{2}}{\partial z^{2}}G^{(0)}(\textbf{r},\textbf{r}^{\prime})\vert_{z=0}+h(\textbf{r}_\parallel)\frac{\partial}{\partial z}G^{(1)}(\textbf{r},\textbf{r}^{\prime})\vert_{z=0}+  G^{(2)}(\textbf{r},\textbf{r}^{\prime})\vert_{z=0}\right\}
\\ 
&+\varepsilon^{3}\left\{[h(\textbf{r}_\parallel)]^{3}\frac{1}{6}\frac{\partial^{3}}{\partial z^{3}}G^{(0)}(\textbf{r},\textbf{r}^{\prime})\vert_{z=0}+[h(\textbf{r}_\parallel)]^{2}\frac{1}{2}\frac{\partial^{2}}{\partial z^{2}}G^{(1)}(\textbf{r},\textbf{r}^{\prime})\vert_{z=0} +h(\textbf{r}_\parallel)\frac{\partial}{\partial z}G^{(2)}(\textbf{r},\textbf{r}^{\prime})\vert_{z=0}+G^{(3)}(\textbf{r},\textbf{r}^{\prime})\vert_{z=0}\right\}+...=0. 
\label{taylor-expandida}
\end{align}
Since $\varepsilon$ is an arbitrary parameter, the solution of Eq. \eqref{taylor-expandida} requires that the coefficients multiplying each power of $\varepsilon$ vanish.
In this way, one obtains boundary conditions at $z=0$ to $G^{(0)}$ and $G^{(n)}$, which are given by \cite{Clinton-PRBII-1985}:
\begin{align}
\nonumber
&G^{(0)}(\textbf{r},\textbf{r}^{\prime})\vert_{z=0} = 0 ,
\\
\nonumber
&G^{(1)}(\textbf{r},\textbf{r}^{\prime})\vert_{z=0} =- h(\textbf{r}_\parallel)\frac{\partial}{\partial z}G^{(0)}(\textbf{r},\textbf{r}^{\prime})\vert_{z=0} ,
\\
\nonumber 
&G^{(2)}(\textbf{r},\textbf{r}^{\prime})\vert_{z=0} =- \left[h(\textbf{r}_\parallel)\right]^{2}\frac{1}{2}\frac{\partial^{2}}{\partial z^{2}}G^{(0)}(\textbf{r},\textbf{r}^{\prime})\vert_{z=0}-h(\textbf{r}_\parallel)\frac{\partial}{\partial z}G^{(1)}(\textbf{r},\textbf{r}^{\prime})\vert_{z=0} ,
\\ 
\nonumber
&G^{(3)}(\textbf{r},\textbf{r}^{\prime})\vert_{z=0} =-
\left[h(\textbf{r}_\parallel)\right]^{3}\frac{1}{6}\frac{\partial^{3}}{\partial z^{3}}G^{(0)}(\textbf{r},\textbf{r}^{\prime})\vert_{z=0}-\left[h(\textbf{r}_\parallel)\right]^{2}\frac{1}{2}\frac{\partial^{2}}{\partial z^{2}}G^{(1)}(\textbf{r},\textbf{r}^{\prime})\vert_{z=0} -h(\textbf{r}_\parallel)\frac{\partial}{\partial z}G^{(2)}(\textbf{r},\textbf{r}^{\prime})\vert_{z=0},\\ 
\nonumber
&... 
\end{align}

\end{widetext}
We can put all these equations together as follows
\begin{align}
\label{G0-Boundary}
G^{(0)}(\textbf{r},\textbf{r}^{\prime})\vert_{z=0} &= 0, 
\\ \nonumber
G^{(n)}(\textbf{r},\textbf{r}^{\prime})\vert_{z=0}  &=
\\ \label{Gn-Boundary}
-\sum_{m=1}^{n}&\frac{[h(\textbf{r}_\parallel)]^{m}}{m!}\frac{\partial^{m}}{\partial z^{m}}G^{(n-m)}(\textbf{r},\textbf{r}^{\prime})\vert_{z=0}\quad (n\geq 1).
\end{align}
From these equations, it is important to remark that only by knowing $G^{(0)}$, one can obtain the boundary condition for $G^{(1)}$, and then for $G^{(2)}$, $G^{(3)}$, and so on.
Also note that for $n \geq 1$ these boundary conditions are inhomogeneous, and thus, the problem of find each perturbative correction for the corrugated situation requires to solve the homogeneous equation given by Eq. \eqref{poisson-ordem-n}, with inhomogeneous boundary conditions, considered at $z=0$, given by Eq. \eqref{Gn-Boundary}.
In summary, the solution for the unperturbed problem, related to the case of a plane surface, can be obtained by solving the following set of equations:
\begin{align}
	\boldsymbol{\nabla}^2 G^{(0)}(\textbf{r},\textbf{r}^{\prime})
	&=-4\pi \delta\left({\bf r}-{\bf r}^{\prime}\right),
	\label{poisson-ordem-0-sumarizando}
	\\
	G^{(0)}(\textbf{r},\textbf{r}^{\prime})\vert_{z=0} &= 0,
	\label{G0-Boundary-sumarizando}\\
	G^{(0)}\left.\left(\textbf{r},\textbf{r}^{\prime}\right)\right|_{|\textbf{r}|\gg |\textbf{r}^\prime|}&\to 0,
	\label{G0-bc-infinity}
\end{align}   
and when this solution is obtained, one can solve, for $n\geq 1$, the following set of equations, related to the case of a corrugated surface:
\begin{align}
\boldsymbol{\nabla}^2 G^{(n)}(\textbf{r},\textbf{r}^{\prime})&=  0,
\label{poisson-ordem-n-sumarizando}
\\
G^{(n)}(\textbf{r},\textbf{r}^{\prime})\vert_{z=0} &= -\sum_{m=1}^{n}\frac{[h(\textbf{r}_\parallel)]^{m}}{m!}\frac{\partial^{m}}{\partial z^{m}}G^{(n-m)}(\textbf{r},\textbf{r}^{\prime})|_{z=0},
\label{Gn-Boundary-sumarizando}
\\
G^{(n)}(\textbf{r},\textbf{r}^{\prime})|_{|\textbf{r}|\gg |\textbf{r}^\prime|}&\to0.
\label{Gn-bc-infinity}
\end{align}   
Note that the original problem of solving the non-homogeneous Eq. \eqref{poisson}, under a homogeneous boundary condition on a complicated surface $z=h(\textbf{r}_\parallel)$ [Eq. \eqref{bc-corrugated}], is now replaced by the problem of solving the homogeneous Eq. \eqref{poisson-ordem-n-sumarizando}, under non-homogeneous boundary conditions on a simple plane at $z=0$ [Eq. \eqref{Gn-Boundary-sumarizando}].
Following Ref. \cite{Clinton-PRBII-1985}, in order to find a solution for Eqs. \eqref{poisson-ordem-0-sumarizando}-\eqref{Gn-bc-infinity}, it is convenient to introduce the Fourier transformation of a function $f(\textbf{r},\textbf{r}^{\prime})$ as
\begin{eqnarray}
\tilde{f}\left(\textbf{k},\textbf{r}_{\parallel}^{\prime};z,z^{\prime}\right)=\int d^{2}\textbf{r}_{\parallel}e^{-i\textbf{k}\cdot\textbf{r}_{\parallel}}f(\textbf{r},\textbf{r}^{\prime}),
\label{Fourier}
\end{eqnarray} 
and its inverse transformation as
\begin{eqnarray}
	f(\textbf{r},\textbf{r}^{\prime})=\frac{1}{\left(2\pi\right)^{2}}\int d^{2}\textbf{k}e^{i\textbf{k}\cdot\textbf{r}_{\parallel}}
	\tilde{f}\left(\textbf{k},\textbf{r}_{\parallel}^{\prime};z,z^{\prime}\right).
	\label{Inverse-Fourier}
\end{eqnarray} 
In this way, by writing $G(\textbf{r},\textbf{r}^{\prime})$, $G^{(0)}(\textbf{r},\textbf{r}^{\prime})$ and $G^{(n)}(\textbf{r},\textbf{r}^{\prime})$ as an inverse Fourier transformation, Eq. \eqref{Green_Gen} can be written as
\begin{align}
\tilde{G}\left(\textbf{k},\textbf{r}_{\parallel}^{\prime};z,z^{\prime}\right)&= \tilde{G}^{(0)}\left(\textbf{k},\textbf{r}_{\parallel}^{\prime};z,z^{\prime}\right)
\nonumber
\\
& +\sum_{n=1}^{\infty}\varepsilon^n\tilde{G}^{(n)}\left(\textbf{k},\textbf{r}_{\parallel}^{\prime};z,z^{\prime}\right),
\label{G-til}
\end{align}
and, by knowing that the Dirac delta function is given in the Fourier space as
\begin{align}
\delta\left(\textbf{r}-\textbf{r}^{\prime}\right)&=\frac{1}{\left(2\pi\right)^{2}}\int d^{2}\textbf{k}e^{i\textbf{k}\cdot\left(\textbf{r}_{\parallel}-\textbf{r}_{\parallel}^{\prime}\right)}\delta\left(z-z^{\prime}\right), \label{delta}
\end{align}
we can also write Eqs. \eqref{poisson-ordem-0-sumarizando} and \eqref{poisson-ordem-n-sumarizando} as \cite{Clinton-PRBII-1985}:
\begin{align}
	\label{G0}
	&\left(\frac{\partial^{2}}{\partial z^{2}}-\left|\textbf{k}\right|^{2}\right)\tilde{G}^{(0)}\left(\textbf{k},\textbf{r}_{\parallel}^{\prime};z,z^{\prime}\right) = -4\pi e^{-i\textbf{k}\cdot\textbf{r}_{\parallel}^{\prime}}\delta\left(z-z^{\prime}\right), \\
	\label{Gn}
	&\left(\frac{\partial^{2}}{\partial z^{2}}-\left|\textbf{k}\right|^{2}\right)\tilde{G}^{(n)}\left(\textbf{k},\textbf{r}_{\parallel}^{\prime};z,z^{\prime}\right) = 0\;\;(n\geq1).
\end{align} 
Besides this, one can also find the boundary conditions given by Eqs. \eqref{G0-Boundary-sumarizando} and \eqref{Gn-Boundary-sumarizando} in the Fourier space \cite{Clinton-PRBII-1985} (see Appendix \ref{transf-of-gn}):
\begin{align}
	\label{G0-Boundary-Fourier}
	\tilde{G}^{(0)}\left(\textbf{k},\textbf{r}_{\parallel}^{\prime};z,z^{\prime}\right)\vert_{z=0} = & 0, 
	\\
	\tilde{G}^{(n)}(\textbf{k},\textbf{r}_{\parallel}^{\prime};z,z^{\prime})\vert_{z=0} = & -\sum_{m=1}^{n}\int\frac{d^{2}\textbf{k}^{\prime}}{(2\pi)^{2}}\frac{\tilde{h}_{m}(\textbf{k}-\textbf{k}^{\prime})}{m!}
	\nonumber \\ \label{Gn-Boundary-Fourier}
	 &\times\frac{\partial^{m}}{\partial z^{m}}\tilde{G}^{(n-m)}(\textbf{k}^{\prime},\textbf{r}_{\parallel}^{\prime};z,z^{\prime})\vert_{z=0},
\end{align} 
where
\begin{equation}
\label{h-k}
\tilde{h}_{m}(\textbf{k}-\textbf{k}^{\prime})=\int d^{2}\textbf{r}_{\parallel}e^{-i(\textbf{k}-\textbf{k}^{\prime})\cdot \textbf{r}_{\parallel}}\left[h(\textbf{r}_\parallel)\right]^{m}.
\end{equation}

Under the boundary condition given by Eq. \eqref{G0-Boundary-Fourier} and the condition that $\tilde{G}^{(0)}$ goes to zero for large distances [Eq. \eqref{G0-bc-infinity}], one can obtain that the solution of Eq. \eqref{G0} is given by (see Appendix \ref{calc-of-g0})
\begin{equation}
\label{Solu-G0-Fourier}
\tilde{G}^{(0)}\left(\textbf{k},\textbf{r}_{\parallel}^{\prime};z,z^{\prime}\right)=\frac{2\pi}{\left|\textbf{k}\right|}e^{-i\textbf{k}\cdot\textbf{r}_{\parallel}^{\prime}} \left[e^{-\left|\textbf{k}\right|\left|z-z^{\prime}\right|}-e^{-\left|\textbf{k}\right|\left(z+z^{\prime}\right)}\right].
\end{equation}
By performing an inverse Fourier transformation on $\tilde{G}^{(0)}$, one obtains
\begin{eqnarray}
\nonumber
G^{(0)}(\textbf{r},\textbf{r}^{\prime})&=&\frac{1}{\left[|\textbf{r}_{\parallel}-\textbf{r}_{\parallel}^{\prime}|^{2}+\left(z-z^{\prime}\right)^{2}\right]^{\frac{1}{2}}}
\\
&-&\frac{1}{\left[|\textbf{r}_{\parallel}-\textbf{r}_{\parallel}^{\prime}|^{2}+\left(z+z^{\prime}\right)^{2}
\right]^{\frac{1}{2}}}, \label{Solu-G0}
\end{eqnarray}
which is the solution already obtained via image method and shown in Eq. \eqref{phi0}.

Under the boundary condition given by Eq. \eqref{Gn-Boundary-Fourier} and the condition that $\tilde{G}^{(n)}$ goes to zero for large distances [Eq. \eqref{Gn-bc-infinity}], one can obtain that the solution of Eq. \eqref{Gn} is given by (see Appendix \ref{calc-of-gn})
\begin{align}
\nonumber
\tilde{G}^{(n)}\left(\textbf{k},\textbf{r}_{\parallel}^{\prime};z,z^{\prime}\right) &=  -e^{-\left|\textbf{k}\right|z}\sum_{m=1}^{n}\int\frac{d^{2}\textbf{k}^{\prime}}{(2\pi)^{2}}\frac{\tilde{h}_{m}(\textbf{k}-\textbf{k}^{\prime})}{m!}
\\ \label{Solu-Gn-Fourier}
&\times\frac{\partial^{m}}{\partial z^{m}}\tilde{G}^{(n-m)}(\textbf{k}^{\prime},\textbf{r}_{\parallel}^{\prime};z,z^{\prime})\vert_{z=0}. 
\end{align}
From this equation, one obtains that
the functions $\tilde{G}^{(n)}$
are given recursively in terms of $\tilde{G}^{(0)}$.
The solution for ${G}^{(n)}$ is obtained by performing the inverse Fourier transformation,
\begin{eqnarray}
	G^{(n)}(\textbf{r},\textbf{r}^{\prime})=\frac{1}{\left(2\pi\right)^{2}}\int d^{2}\textbf{k}e^{i\textbf{k}\cdot\textbf{r}_{\parallel}}
	\tilde{G}^{(n)}\left(\textbf{k},\textbf{r}_{\parallel}^{\prime};z,z^{\prime}\right).
	\label{Solu-Gn}
\end{eqnarray} 
with $\tilde{G}^{(n)}$ given by Eq. \eqref{Solu-Gn-Fourier}.
Thus, substituting Eqs. \eqref{Solu-G0} and \eqref{Solu-Gn} in Eq. \eqref{Green_Gen} (with $\varepsilon=1$), one obtains the Green's function related to the problem of a point charge in the presence of a grounded conducting corrugated surface described by $z=h(\textbf{r}_\parallel)$.
In the next two sections we focus on the application of this approach to the cases of a slightly translated plane and of a sinusoidal corrugated surface.

\section{Point charge in the presence of a slightly translated plane}
\label{translated-surface}

In this section, for pedagogical purposes, we perform a consistency check of the perturbative formulas obtained in the previous section, by applying them to the case of a point charge $Q$ in the presence of a slightly translated plane, whose exact Green's function is known and with which we can compare the perturbative results. 
We also use this case to discuss some conditions under which a perturbative solution becomes closer to the exact one.
\begin{figure}
\centering  
\subfigure[\label{plano-deslocado-cima}]{\label{plano-deslocado-cima}\epsfig{file=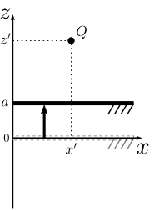, width=0.4 \linewidth}}
\subfigure[\label{plano-deslocado-baixo}]{\label{plano-deslocado-baixo}\epsfig{file=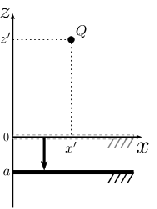, width=0.4 \linewidth}}
\caption{
Cross section view of a charge $Q$, located at ${\bf r}^{\prime}={\bf r}_{\parallel}^{\prime}+z^\prime\hat{{\bf z}}$ (with $z^\prime>0$), in the presence of a grounded conducting plane, translated from $z=0$ to $z=a$.
In (a) we have $a>0$, and 
in (b) we have $a<0$.
}
\label{plano-deslocado}
\end{figure}

To apply the perturbative formulas discussed in Sec. \ref{review-clinton} to the case
of a point charge $Q$ in the presence of a slightly translated plane 
(from $z=0$ to $z=a$, as illustrated in Fig. \ref{plano-deslocado}),
we consider in these formulas $h\left(\textbf{r}_{\parallel}\right)=a$, with $|a|<z^\prime$.
In this way, from Eq. \eqref{h-k}, we have
\begin{equation}
\tilde{h}_{m}(\textbf{k}-\textbf{k}^{\prime})=a^{m}
\left(2\pi\right)^{2}\delta\left(\textbf{k}-\textbf{k}^{\prime}\right),
\end{equation}
so that, from Eq. \eqref{Solu-Gn-Fourier}, one obtains
(up to the third order of perturbation):
\begin{align}
\tilde{G}^{\left(1\right)}\left(\textbf{k},\textbf{r}_{\parallel}^{\prime};z,z^{\prime}\right) & =-4\pi ae^{-i\textbf{k}\cdot\textbf{r}_{\parallel}^{\prime}}e^{-\left|\textbf{k}\right|\left(z+z^{\prime}\right)},\\
\tilde{G}^{\left(2\right)}\left(\textbf{k},\textbf{r}_{\parallel}^{\prime};z,z^{\prime}\right) & =-4\pi a^{2}\left|\textbf{k}\right|e^{-i\textbf{k}\cdot\textbf{r}_{\parallel}^{\prime}}e^{-\left|\textbf{k}\right|\left(z+z^{\prime}\right)},\\
\tilde{G}^{\left(3\right)}\left(\textbf{k},\textbf{r}_{\parallel}^{\prime};z,z^{\prime}\right) & =-\frac{8}{3}\pi a^{3}\left|\textbf{k}\right|^{2}e^{-i\textbf{k}\cdot\textbf{r}_{\parallel}^{\prime}}e^{-\left|\textbf{k}\right|\left(z+z^{\prime}\right)}.
\end{align}
By performing the inverse Fourier transformation, one obtains
\begin{align}
\label{g1-a}
G^{\left(1\right)}\left(\textbf{r},\textbf{r}^{\prime}\right) & =-\frac{2\left(z+z^{\prime}\right)}{\left[|\textbf{r}_{\parallel}-\textbf{r}_{\parallel}^{\prime}|^{2}+\left(z+z^{\prime}\right)^{2}\right]^{3/2}}a,\\
\label{g2-a}
G^{\left(2\right)}\left(\textbf{r},\textbf{r}^{\prime}\right) & =-\frac{2\left[2\left(z+z^{\prime}\right)^{2}-|\textbf{r}_{\parallel}-\textbf{r}_{\parallel}^{\prime}|^{2}\right]}{\left[|\textbf{r}_{\parallel}-\textbf{r}_{\parallel}^{\prime}|^{2}+\left(z+z^{\prime}\right)^{2}\right]^{5/2}}a^{2},\\
\label{g3-a}
G^{\left(3\right)}\left(\textbf{r},\textbf{r}^{\prime}\right) & =-\frac{4\left(z+z^{\prime}\right)\left[2\left(z+z^{\prime}\right)^{2}-3|\textbf{r}_{\parallel}-\textbf{r}_{\parallel}^{\prime}|^{2}\right]}{\left[|\textbf{r}_{\parallel}-\textbf{r}_{\parallel}^{\prime}|^{2}+\left(z+z^{\prime}\right)^{2}\right]^{7/2}}a^{3}.
\end{align}
Thus, from Eq. \eqref{Green_Gen} (with $\varepsilon=1$), the Green's function for this problem, up to the third perturbative order, is given by $G\approx G^{\left(0\right)} + G^{\left(1\right)} + G^{\left(2\right)} + G^{\left(3\right)}$, with $G^{\left(s\right)}$ $(s=0,1,2,3)$ given by
Eqs. \eqref{Solu-G0}, \eqref{g1-a}, \eqref{g2-a}, and \eqref{g3-a}.

The exact result, for the same problem of a translated plane, 
can be obtained by means of the image method, as shown in Sec. \ref{review-point-plane},
so that one finds
\begin{eqnarray}
\nonumber
G\left(\textbf{r},\textbf{r}^{\prime}\right)&=&\frac{1}{\left[|\textbf{r}_{\parallel}-\textbf{r}_{\parallel}^{\prime}|^{2}+\left|z-z^{\prime}\right|^{2}\right]^{\frac{1}{2}}}
\\
&-&\frac{1}{\left[|\textbf{r}_{\parallel}-\textbf{r}_{\parallel}^{\prime}|^{2}+\left(z+z^{\prime}-2a\right)^{2}\right]^{\frac{1}{2}}}, \label{g0-a-exato}
\end{eqnarray}
which, for $a=0$, recovers Eq. \eqref{phi0}.
By expanding Eq. \eqref{g0-a-exato} in powers of $a$, one obtains
\begin{eqnarray}
\label{g0-a-expanded}
G(\textbf{r},\textbf{r}^{\prime})&=&G^{(0)}(\textbf{r},\textbf{r}^{\prime})
-\frac{2\left(z+z^{\prime}\right)}{\left[|\textbf{r}_{\parallel}-\textbf{r}_{\parallel}^{\prime}|^{2}+\left(z+z^{\prime}\right)^{2}\right]^{\frac{3}{2}}}a
\nonumber\\
&-&\frac{2\left[2\left(z+z^{\prime}\right)^{2}-|\textbf{r}_{\parallel}-\textbf{r}_{\parallel}^{\prime}|^{2}\right]}{\left[|\textbf{r}_{\parallel}-\textbf{r}_{\parallel}^{\prime}|^{2}+\left(z+z^{\prime}\right)^{2}\right]^{\frac{5}{2}}}a^{2}
\nonumber\\
&-&\frac{4\left(z+z^{\prime}\right)\left[2\left(z+z^{\prime}\right)^{2}-3|\textbf{r}_{\parallel}-\textbf{r}_{\parallel}^{\prime}|^{2}\right]}{\left[|\textbf{r}_{\parallel}-\textbf{r}_{\parallel}^{\prime}|^{2}+\left(z+z^{\prime}\right)^{2}\right]^{\frac{7}{2}}}a^{3}
\nonumber\\
&+&...,
\end{eqnarray}
where $G^{(0)}(\textbf{r},\textbf{r}^{\prime})$ is given by Eq. \eqref{Solu-G0}.
Note that, as expected for this consistency check, the terms of this expansion, proportional
to $a$, $a^2$ and $a^3$, coincide, respectively, with those 
shown in Eqs. \eqref{g1-a}, \eqref{g2-a} and \eqref{g3-a},
which were obtained by means of the perturbative approach discussed in Sec. \ref{review-clinton}.

\begin{figure}
\centering  
\epsfig{file=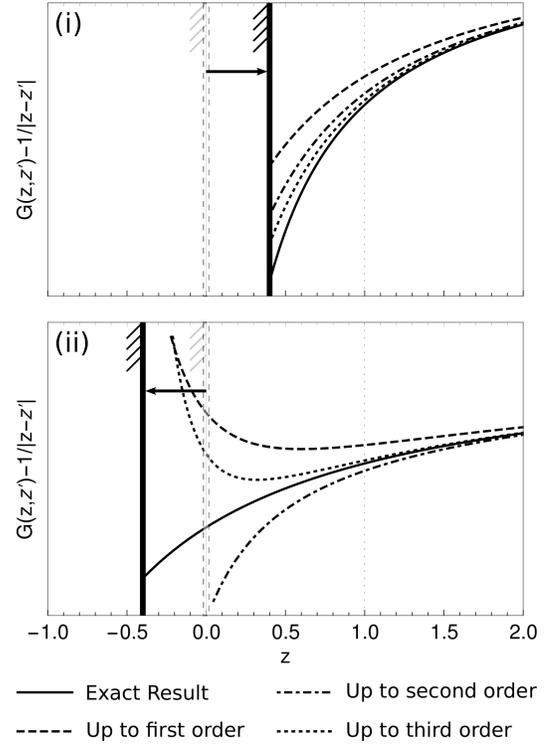, width=.8 \linewidth}
\caption{
Comparison between the exact solution for the Green's function for the case of
a translated plane $(z=a)$, given by Eq. \eqref{g0-a-exato}, and approximate solutions
up to first, second and third perturbative order.
We consider $|\textbf{r}_{\parallel}-\textbf{r}_{\parallel}^{\prime}| = 0$, and $z^\prime=1$. 
In (i), we show the case where $a=0.4$, whereas in (ii) it is shown  the case with $a=-0.4$.
}
\label{fig:convergencia-plano-deslocado}
\end{figure}

Using this simple case of a translated plane, we can also discuss some conditions under which a perturbative solution becomes closer to the exact one.
Considering, for simplicity, $|\textbf{r}_{\parallel}-\textbf{r}_{\parallel}^{\prime}| = 0$, in Fig. \ref{fig:convergencia-plano-deslocado} we compare the exact solution given by Eq. \eqref{g0-a-exato} with the approximate ones up to first, second and third perturbative order.
Note that an approximate result, up to a given order, 
becomes closer to the exact one as $z$ increases,
and further as $z$ approaches to $a$.
When comparing the case where $a>0$ [Fig. \ref{fig:convergencia-plano-deslocado}(i)] with that where $a<0$ [Fig. \ref{fig:convergencia-plano-deslocado}(ii)], one can see that the perturbative results are closer to the exact one in (i) than in (ii), specially in the region close to the surface.
In this way, if we are interested in calculating physical quantities near or on the surface,
as the electric field and induced surface charge density, the consideration of $a>0$ 
leads to a better approximate solution.
Extending these ideas for a general corrugation, hereafter we will focus on 
models with $h(\textbf{r}_\parallel) \geq 0$.
%
%
%

\section{Point charge in the presence of a sinusoidal conducting surface}
\label{sinusoidal-surface}

\begin{figure}
\centering
\epsfig{file=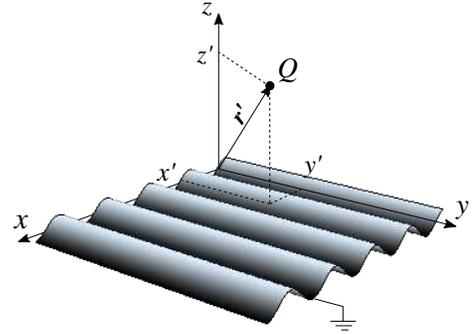,  width=0.7 \linewidth}
\caption{{
Illustration of a charge $Q$, located at ${\bf r}^{\prime}={\bf r}_{\parallel}^{\prime}+z^\prime\hat{{\bf z}}$ (with $z^\prime>0$), interacting with a grounded conducting sinusoidal surface, whose corrugation profile is described by $h(\textbf{r}_\parallel)=a[1+\cos\left(\nu x\right)]/2$.
}}
\label{fig:carga-superficie-senoid}
\end{figure}

As another application of the Clinton, Esrick and Sacks calculations, let us study the image potential, calculated up to the first perturbative order, for the problem of a point charge $Q$ in the presence of a sinusoidal corrugated surface, described by (see Fig. \ref{fig:carga-superficie-senoid})
\begin{equation}
h(\textbf{r}_\parallel)=\frac{a}{2}[1+\cos\left(\nu x\right)]. \label{h-seno}
\end{equation}

The approximate solution for $G$, up to first-order in $h$, is given by
\begin{equation}
G(\textbf{r},\textbf{r}^\prime)\approx G^{(0)}(\textbf{r},\textbf{r}^\prime)+G^{(1)}(\textbf{r},\textbf{r}^\prime),
\end{equation}
with $G^{(0)}$ given by Eq. \eqref{Solu-G0}.
To calculate $G^{(1)}$, we first substitute Eq. \eqref{h-seno} in Eq. \eqref{h-k} (for $m=1$), and obtain that
\begin{align}
	\nonumber
	\tilde{h}_{1}(\mathbf{k}-\mathbf{k^{\prime}}) & =a\pi^{2}\delta(k_{y}-k_{y}^{\prime})[2\delta(k_{x}-k_{x}^{\prime})\\
	&+\delta(k_{x}-k_{x}^{\prime}+\nu) +\delta(-k_{x}+k_{x}^{\prime}+\nu)].
\end{align}
Substituting it in Eq. \eqref{Solu-Gn-Fourier}, for $n=1$, one obtains
\begin{align}
	\tilde{G}^{(1)}\left(\mathbf{k},\mathbf{r^\prime_\parallel};z,z^\prime\right)&=-\pi a e^{-i\mathbf{k}\cdot\mathbf{r^\prime_\parallel}-kz}\left[e^{i\nu x^\prime-z^\prime\sqrt{(k_x-\nu)^2+k_y^2}}\right.\nonumber\\
	&\left.+e^{-i\nu x^\prime-z^\prime\sqrt{(k_x+\nu)^2+k_y^2}}\right],
\end{align}
and $G^{(1)}(\textbf{r},\textbf{r}^\prime)$ is obtained by performing an inverse Fourier transformation of this equation.

\begin{figure}
\centering  
\subfigure[\label{fig:linhas-campo-plano}]{\epsfig{file=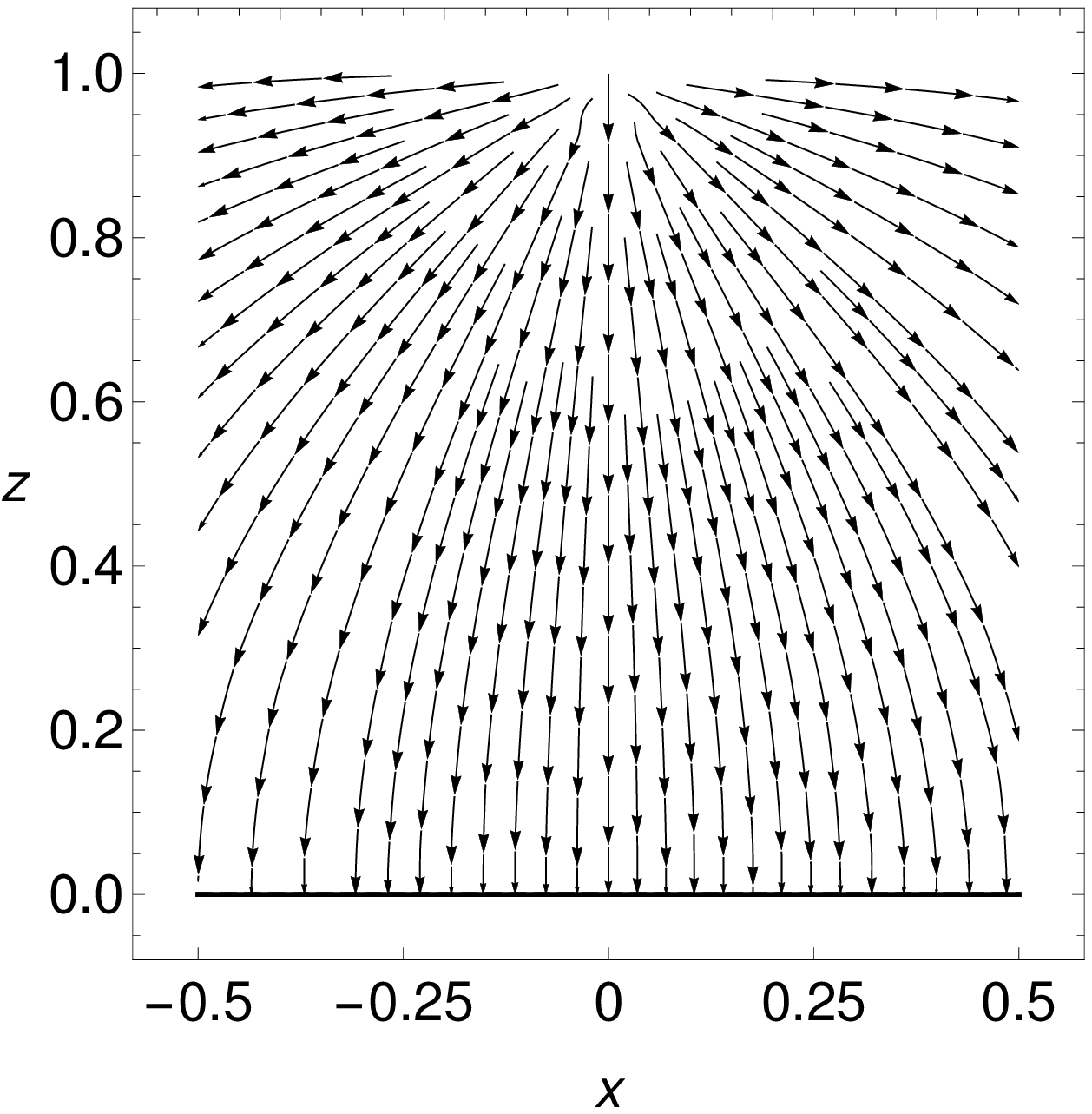, width=0.7 \linewidth}}
\hspace{2mm}
\subfigure[\label{fig:linhas-campo-senoidal}]{\epsfig{file=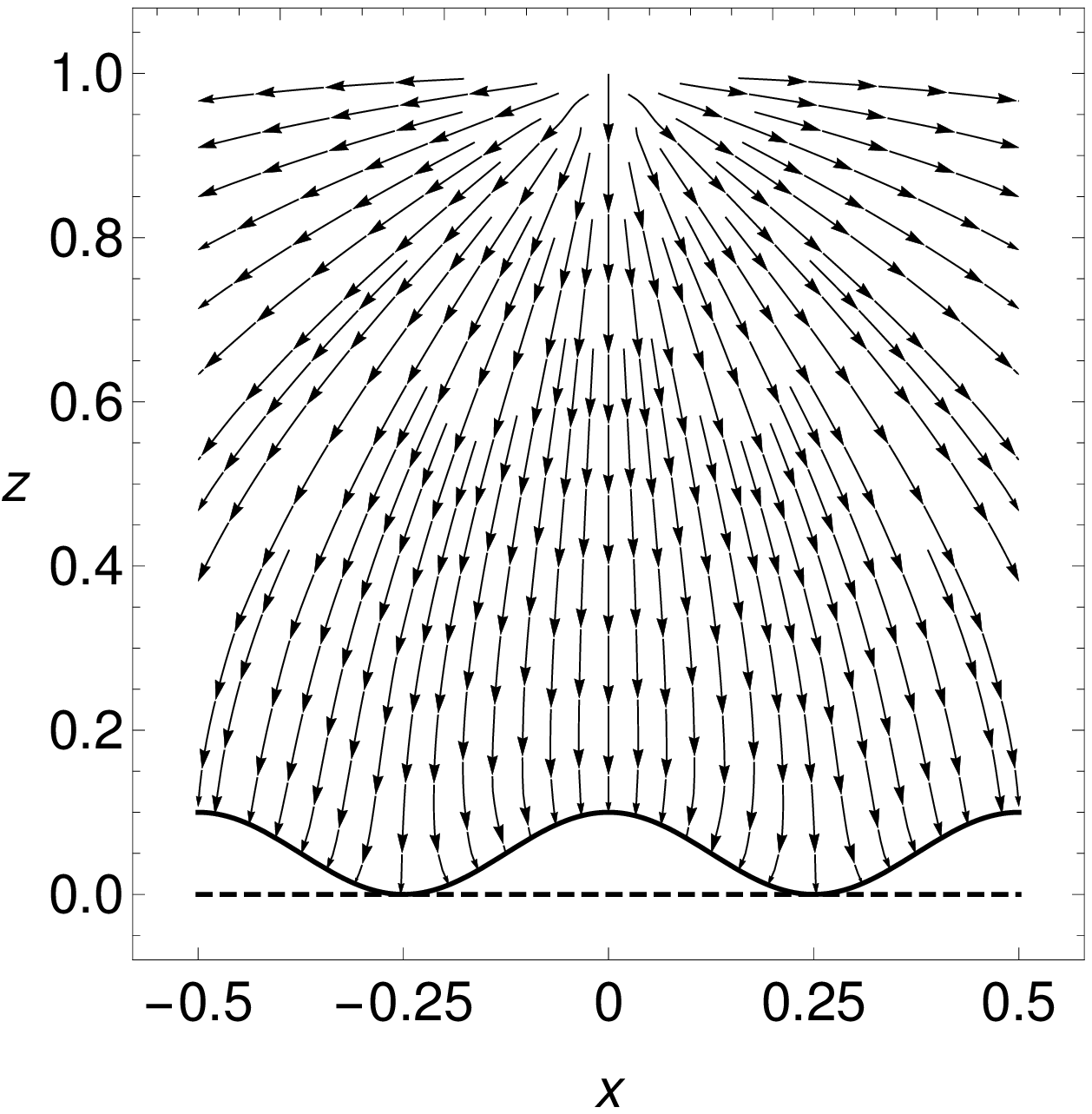, width=0.7 \linewidth}}
\caption{
Behavior of the electric field lines, along the plane $y=0$, generated by a point charge $(Q=1)$ at $\textbf{r}^\prime=\hat{\textbf{z}}$ in the presence of: (a) a plane surface; (b) a sinusoidal corrugated surface with $a=0.1$ and $\nu=4\pi$.
}
\label{fig:linhas-campo}
\end{figure}

As a first application of the above approximate solution,
we calculate the electric field $\textbf{E}$, which is shown in 
Fig. \ref{fig:linhas-campo}.
In Fig. \ref{fig:linhas-campo}(a), we show the electric field lines for the case of 
a point charge in the presence of a planar surface at $z=0$.
In Fig. \ref{fig:linhas-campo}(b), we consider the case of a sinusoidal surface and 
show the electric field lines modified by the corrugation.
In addition, for the considered values for $a$, $\nu$ and $z^\prime$, one
can see that the electric field lines are practically perpendicular to the surface, which shows that, for this case, we have an acceptable approximate solution.

\begin{figure}
\centering  
\epsfig{file=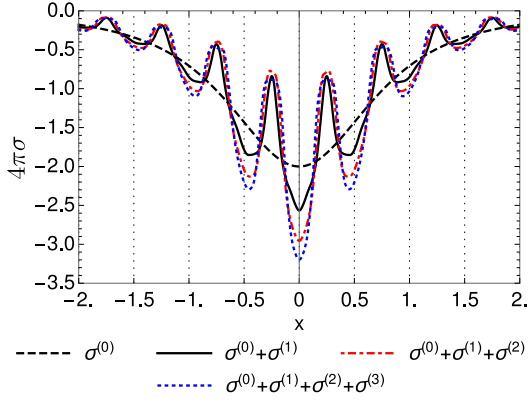, width=0.8 \linewidth}
\caption{
Behavior of the induced surface charge density $\sigma(x,y=0)$, due to the presence of a point charge $(Q=1)$ located at $\textbf{r}^\prime=\hat{\textbf{z}}$.
Here, $\sigma^{(0)}$ (dashed line) represents the case of a plane surface at $z=0$, whereas $\sigma^{(0)}+\sigma^{(1)}$ (solid line), $\sigma^{(0)}+\sigma^{(1)}+\sigma^{(2)}$ (dot-dashed line), and $\sigma^{(0)}+\sigma^{(1)}+\sigma^{(2)}+\sigma^{(3)}$ (dotted line) represent the cases of a sinusoidal corrugated surface, with $a=0.1$ and $\nu=4\pi$, calculated up to first, second and third perturbative order, respectively.
}
\label{fig:densidade-carga}
\end{figure}

As another application, we calculate the induced surface charge density
$\sigma$, which is given by (see, for instance, Ref. \cite{Griffiths})
\begin{equation}
	\label{densidade-carga}
	\sigma(\textbf{r}_\parallel) = \frac{1}{4\pi}[\textbf{E} \cdot \hat{{\bf n}}]_{z\to h(\textbf{r}_\parallel)},
\end{equation}
where $\hat{{\bf n}}$ is a unit vector normal to the surface.
In Fig. \ref{fig:densidade-carga}, we show how the charge density is
changed when considering a sinusoidal surface.
Note that, due to the oscillatory profile of the surface, the values of the induced surface charge density for this case oscillates around those for a plane surface.
We also compute the corrections coming from the second and third perturbative orders,
and show that, for the considered values for $a$, $\nu$ and $z^\prime$, these corrections do not change the qualitative behavior already obtained considering only the first perturbative order.

\section{Conducting cylinder placed in a uniform electric field}
\label{normal-cylinder}

A cylinder in the presence of a uniform electrostatic field is another very 
common exercise in introductory level electromagnetism courses 
(see, for instance, Refs. \cite{Griffiths, Machado,Milford,Jackson,Nayfeh-Electrodynamics-1985,Arfken}). 
Therefore, in this section we make a brief review of this problem to, in the next section, introduce corrugation to the cylinder surface. 

\begin{figure}
\centering
\epsfig{file=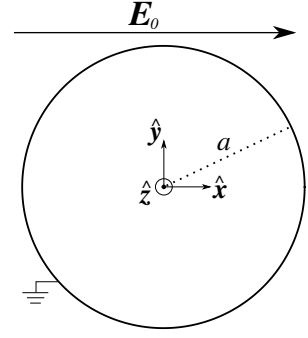,  width=0.45 \linewidth}
\caption{{
Cross section view of an infinite grounded conducting cylinder, with radius $a$, placed in a uniform electrostatic field $\textbf{E}_{0}=E_0 \hat{{\bf x}}$.
}}
\label{cilindro-liso}
\end{figure}

We start by considering an infinite grounded conducting cylinder, with radius $a$, placed in a uniform electrostatic field $\textbf{E}_{0}=E_0 \hat{{\bf x}}$, as illustrated in Fig. \ref{cilindro-liso}. 
The electrostatic potential for this case can be calculated by means of the Laplace's equation,
\begin{eqnarray}
	\nabla^{2}\Phi\left({\bf r},{\bf r}^{\prime}\right)=0,
	\label{laplace-equation}
\end{eqnarray}
which, in cylindrical coordinates, can be written as
\begin{align}
	\nonumber
	\frac{1}{\rho}\frac{\partial}{\partial\rho}\left(\rho\frac{\partial}{\partial\rho}\right)\Phi\left(\rho,\theta,z\right)+&\frac{1}{\rho^{2}}\frac{\partial^{2}}{\partial\theta^{2}}\Phi\left(\rho,\theta,z\right)
	\\
	&+\frac{\partial^{2}}{\partial z^{2}}\Phi\left(\rho,\theta,z\right)=0.
\end{align}
Since the cylinder is considered to be very long, we have a symmetry in the $z$-axis, so that $\Phi\left(\rho,\theta,z\right)\to \Phi\left(\rho,\theta\right)$, and the Laplace's equation can be written as
\begin{equation}
	\frac{1}{\rho}\frac{\partial}{\partial\rho}\left(\rho\frac{\partial}{\partial\rho}\right)\Phi\left(\rho,\theta\right)+\frac{1}{\rho^{2}}\frac{\partial^{2}}{\partial\theta^{2}}\Phi\left(\rho,\theta\right)=0.
	\label{laplace_cili}
\end{equation}
By using the method of separation of variables, we obtain \cite{Arfken}
\begin{align}
\nonumber
\Phi\left(\rho,\theta\right)=&\sum_{m=1}^{\infty} \left(A_{m}\rho^{m}+B_{m}\rho^{-m}\right)
\\ 
&\times\left[C_{m}\cos\left(m\theta\right)+D_{m}\sin\left(m\theta\right)\right].
\label{solu_lap_cili}
\end{align}
Since we consider the cylinder as a grounded perfect conductor, we have a boundary condition at $\rho=a$, which is given by
\begin{eqnarray}
\Phi\left(\rho,\theta\right)\vert_{\rho=a}=0.
\label{bc-surf-cyl}
\end{eqnarray}
Besides this, at a large distance $\rho \gg a$, the potential does not go to zero, since we have the presence of the external uniform electric field.
In this case, the potential has to describe the field $\textbf{E}_{0}=E_0 \hat{{\bf x}}$ and, therefore, it can be written as
\begin{eqnarray}
\left.\Phi\left(\rho,\theta\right)\right|_{\rho\gg a}\to-E_{0}\rho\cos\theta.
\label{bc-far-cyl}
\end{eqnarray}
Using, in Eq. \eqref{solu_lap_cili}, the boundary condition given by Eq. \eqref{bc-far-cyl}, one obtains
\begin{align}
-E_{0}\rho\cos\theta=\sum_{m=1}A_{m}\rho^{m}\left[C_{m}\cos\left(m\theta\right)+D_{m}\sin\left(m\theta\right)\right].
\end{align}
By comparing both sides of this equation, one can see that $D_m$ has to be zero, and thus
\begin{eqnarray}
-E_{0}\rho\cos\theta=\sum_{m=1}A_{m}^\prime\rho^{m}\cos\left(m\theta\right),
\end{eqnarray}
where we made $A_{m} C_{m} = A_{m}^\prime$.
By comparing both sides of this equation, we can see that $A_{1}^\prime=-E_{0}$ and $A_{m\geq2}^\prime=0$. 
Therefore, Eq. \eqref{solu_lap_cili} can be written as
\begin{eqnarray}
\Phi\left(\rho,\theta\right)=-E_{0}\rho\cos\left(\theta\right)+\sum_{m=1}B_{m}^\prime\rho^{-m}\cos\left(m\theta\right).
\label{connection-cyl-1}
\end{eqnarray}
where we made $B_{m} C_{m} = B_{m}^\prime$.
Using the boundary condition given by Eq. \eqref{bc-surf-cyl}, we multiply both sides by $\cos\left(l\theta\right)$ and integrate in $\theta$ from $0$ to $2\pi$, so that one obtains
\begin{align}
\nonumber
-E_{0}a&\int_0^{2\pi}\cos\left(l\theta\right)\cos\left(\theta\right)d\theta
\\
&+\sum_{m=1}B_{m}^\prime a^{-m}\int_0^{2\pi}\cos\left(l\theta\right)\cos\left(m\theta\right)d\theta=0.
\end{align}
Using the orthogonality condition of the cosine function, one obtains
\begin{eqnarray}
B_{m}^\prime=E_{0}a^{m+1}\delta_{1m}.
\label{coef-cyl-1}
\end{eqnarray} 
Thus, substituting Eq. \eqref{coef-cyl-1} in Eq. \eqref{connection-cyl-1}, one obtains
\begin{equation}
\Phi\left(\rho,\theta\right)=-E_{0}\rho\cos\left(\theta\right)\left(1-\frac{a^{2}}{\rho^{2}}\right),
\label{solu-Phi-0}
\end{equation}
which is the solution of Eq. \eqref{laplace_cili}, under the condition that the potential describes the field $\textbf{E}_0$ for large distances [Eq. \eqref{bc-far-cyl}] and the boundary condition given by Eq. \eqref{bc-surf-cyl}.
We remark that this solution only describes the region outside the cylinder.
Inside it, the potential is null, since we consider the cylinder as a grounded perfect conductor.
In the next section, we study how this potential is modified by introducing corrugation on the cylinder surface.

\section{Conducting corrugated cylinder placed in a uniform electric field}
\label{corrugated-cylinder}

\begin{figure}
\centering
\epsfig{file=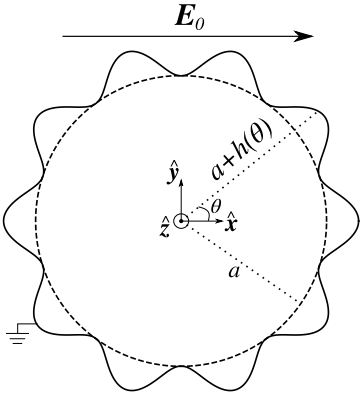,  width=0.55 \linewidth}
\caption{{
Cross section view of an infinite grounded conducting corrugated cylinder (solid line), whose corrugation profile is described by $\rho = a+h(\theta)$, placed in a uniform electrostatic field $\textbf{E}_{0}=E_0 \hat{{\bf x}}$.
}}
\label{cilindro-corrugado}
\end{figure}

We start by considering an infinite grounded conducting cylinder, with radius $a$, placed in a uniform electrostatic field $\textbf{E}_{0}=E_0 \hat{{\bf x}}$, and introduce in it a corrugation described by $\rho = a+h(\theta,z)$, where $h(\theta,z)$ describes a suitable modification $[\text{max}|h(\theta,z)|\ll a]$ on the cylinder surface, as illustrated in Fig. \ref{cilindro-corrugado}. 
Since the cylinder is infinite, we consider, for simplicity, $h(\theta,z)\to h(\theta)$, so that we have a symmetry in the $z$-axis.
Thus, the electrostatic potential in this case can be calculated by means of the Laplace's equation given by Eq. \eqref{laplace_cili}, under the condition that the potential describes the field $\textbf{E}_0$ for large distances [Eq. \eqref{bc-far-cyl}].
Since we consider the cylinder as a grounded perfect conductor, we also have a boundary condition on its surface, which is given by
\begin{equation}
	\Phi(\rho,\theta)\vert_{ \rho = a+h(\theta)}=0.
	\label{bc-cilindro}
\end{equation}

Following the perturbative approach discussed in Sec. \ref{review-clinton},
we introduce an arbitrary auxiliary parameter $\varepsilon$, with $0\leq \varepsilon \leq 1$. 
Thus, we make $h(\theta)\to \varepsilon h(\theta)$, so that
Eq. \eqref{bc-cilindro} is replaced by
\begin{equation}
	\Phi(\rho,\theta)\vert_{ \rho = a+\varepsilon h(\theta)}=0,
	\label{bc-cilindro-epsilon}
\end{equation}
and the electrostatic potential is written as a perturbative expansion in terms of the parameter $\varepsilon$ as
\begin{eqnarray}
	\label{Phi-Gen}
	\Phi\left(\rho,\theta\right)=\Phi^{(0)}\left(\rho,\theta\right)+\sum_{n=1}^{\infty}\varepsilon^{n}\Phi^{(n)}\left(\rho,\theta\right),
\end{eqnarray}
where $\Phi^{(0)}\left(\rho,\theta\right)$ is the solution for the unperturbed problem, related to the case of a non-corrugated cylinder, and the functions $\Phi^{(n)}\left(\rho,\theta\right)$ are the perturbative corrections. 
When $\varepsilon=0$, one has the non-corrugated cylinder $(\rho=a)$, whereas, when $\varepsilon=1$, we recover the actual corrugated cylinder $[\rho=a+h(\theta)]$.
Substituting Eq. \eqref{Phi-Gen} in Eq. \eqref{laplace_cili} one has
\begin{eqnarray}
	\sum_{n=0}^{\infty}
	\varepsilon^{n}\left[\frac{1}{\rho}\frac{\partial}{\partial\rho}\left(\rho\frac{\partial}{\partial\rho}\right)+\frac{1}{\rho^{2}}\frac{\partial^{2}}{\partial\theta^{2}}\right]\Phi^{(n)}\left(\rho,\theta\right)=0.
	\label{eq-laplace-mais-nova}
\end{eqnarray}
Since $\varepsilon$ is an arbitrary parameter, 
the solution of Eq. \eqref{eq-laplace-mais-nova} requires that the coefficients multiplying each
power of $\varepsilon$ vanish.
Then, one has:
\begin{eqnarray}
	\label{Phi0}
	\left[\frac{1}{\rho}\frac{\partial}{\partial\rho}\left(\rho\frac{\partial}{\partial\rho}\right)+\frac{1}{\rho^{2}}\frac{\partial^{2}}{\partial\theta^{2}}\right]\Phi^{(0)}\left(\rho,\theta\right) &=& 0, \\
	\label{Phin}
	\left[\frac{1}{\rho}\frac{\partial}{\partial\rho}\left(\rho\frac{\partial}{\partial\rho}\right)+\frac{1}{\rho^{2}}\frac{\partial^{2}}{\partial\theta^{2}}\right]\Phi^{(n)}\left(\rho,\theta\right) &=& 0\;\;(n\geq1).
\end{eqnarray}
We can find the boundary conditions for Eqs. \eqref{Phi0} and \eqref{Phin} by applying the same procedure discussed in Sec. \ref{review-clinton}. 
Using Eq. \eqref{bc-cilindro-epsilon} in \eqref{Phi-Gen}, and expanding in powers of $\varepsilon h(\theta)$, one has
\begin{eqnarray}
	\sum_{i,m=0}^{\infty}\varepsilon^{i+m}[h\left(\theta\right)]^{m}\frac{1}{m!}\frac{\partial^{m}}{\partial\rho^{m}}\Phi^{(i)}\left(\rho,\theta\right)\vert_{\rho=a}=0.
\end{eqnarray}
Manipulating the series to bring together the terms with a same order in $\varepsilon$, one obtains
\begin{widetext}
\begin{align}\nonumber
	&\Phi^{(0)}\left(\rho,\theta\right)\vert_{\rho=a}+\varepsilon\left[h\left(\theta\right)\frac{\partial}{\partial\rho}\Phi^{(0)}\left(\rho,\theta\right)\vert_{\rho=a}+\Phi^{(1)}\left(\rho,\theta\right)\vert_{\rho=a}\right]
	\\
	&+\varepsilon^{2}\left\{h\left(\theta\right)^{2}\frac{1}{2}\frac{\partial^{2}}{\partial\rho^{2}}\Phi^{(0)}\left(\rho,\theta\right)\vert_{\rho=a}+h\left(\theta\right)\frac{\partial}{\partial\rho}\Phi^{(1)}\left(\rho,\theta\right)\vert_{\rho=a}+\Phi^{(2)}\left(\rho,\theta\right)\vert_{\rho=a}\right\}+...=0.
\end{align}
\end{widetext}
Since $\varepsilon$ is an arbitrary parameter, the solution of this equation requires that the coefficients multiplying each power of $\varepsilon$ vanish.
In this way, we obtain boundary conditions at $\rho=a$ to $\Phi^{(0)}$ and $\Phi^{(n)}$, which are given by:
\begin{align}
\label{Phi0-Boundary}
&\Phi^{(0)}\left(\rho,\theta\right)\vert_{\rho=a} = 0, 
\\ \nonumber
&\Phi^{(n)}\left(\rho,\theta\right)\vert_{\rho=a} =
\\ \label{Phin-Boundary}
&-\sum_{m=1}^{n}\frac{[h\left(\theta\right)]^{m}}{m!}\frac{\partial^{m}}{\partial\rho^{m}}\Phi^{(n-m)}\left(\rho,\theta\right)\vert_{\rho=a}\;\;(n\geq1).
\end{align}
Far away from the cylinder, the modification on the potential due to the presence of corrugation goes to zero, but the potential still has to describe the field $\textbf{E}_0$.
In this way, we also have the following conditions
\begin{align}
\label{Phi0-bc-far}
\Phi^{(0)}\left(\rho,\theta\right)\vert_{\rho \gg a} & \to -E_{0}\rho\cos\theta,
\\
\Phi^{(n)}\left(\rho,\theta\right)\vert_{\rho \gg a} & \to 0.
 \label{Phin-bc-far}
\end{align}

Under the boundary conditions given by Eqs. \eqref{Phi0-Boundary} and \eqref{Phi0-bc-far}, the solution of Eq. \eqref{Phi0} is given by Eq. \eqref{solu-Phi-0}. 
The solution of Eq. \eqref{Phin} can be found by using the method of separation of variables, from which one has
\begin{align} 
	\nonumber
	\Phi^{(n)}\left(\rho,\theta\right)=&\sum_{l=1}^\infty\left(A_{l}\rho^{l}+B_{l}\rho^{-l}\right)
	\\
	&\times\left[C_{l}\cos\left(l\theta\right)+D_{l}\sin\left(l\theta\right)\right].
	\label{step1-solu-cyl-rug}
\end{align}
From Eq. \eqref{Phin-bc-far}, one can conclude that $A_{l}=0$. 
Thus, Eq. \eqref{step1-solu-cyl-rug} becomes
\begin{equation}
	\Phi^{(n)}\left(\rho,\theta\right)=\sum_{l=1}\rho^{-l}\left[C_{l}^\prime\cos\left(l\theta\right)+D_{l}^\prime\sin\left(l\theta\right)\right], \label{usada}
\end{equation}
where we made $B_{l}C_{l}=C_{l}^\prime$, and $B_{l}D_{l}=D_{l}^\prime$.
Using the boundary condition given by Eq. \eqref{Phin-Boundary} in Eq. \eqref{usada}, one has
\begin{equation}
	\Phi^{(n)}\left(\rho,\theta\right)\vert_{\rho=a}=\sum_{l=1}a^{-l}\left[C_{l}^\prime\cos\left(l\theta\right)+D_{l}^\prime\sin\left(l\theta\right)\right].
\end{equation}
Multiplying both sides by $\cos\left(p\theta\right)$ or $\sin\left(p\theta\right)$, and integrating in $\theta$ from $0$ to $2\pi$, one obtains 
\begin{align}
	C_{l}^\prime&=\frac{a^{l}}{\pi}\int_{0}^{2\pi}d\theta\cos\left(l\theta\right)\Phi^{(n)}\left(\rho,\theta\right)\vert_{\rho=a}, \label{A-coef}
	\\
	D_{l}^\prime&=\frac{a^{l}}{\pi}\int_{0}^{2\pi}d\theta\sin\left(l\theta\right)\Phi^{(n)}\left(\rho,\theta\right)\vert_{\rho=a}. \label{B-coef}
\end{align}
Using Eqs. \eqref{A-coef} and \eqref{B-coef} in Eq. \eqref{usada}, one finds
\begin{align}
	\nonumber
	&\Phi^{(n)}\left(\rho,\theta\right) = -\sum_{m=1}^{n}\sum_{l=1}^{\infty}\frac{1}{\pi}\left(\frac{a}{\rho}\right)^{l}
	\\	\label{Solu-Phin}
	&\times\int_{0}^{2\pi}d\bar{\theta}\cos[l(\bar{\theta}-\theta)]\frac{[h(\bar{\theta})]^{m}}{m!}\frac{\partial^{m}}{\partial\rho^{m}}\Phi^{(n-m)}\left(\rho,\bar{\theta}\right)\vert_{\rho=a}.
\end{align}  
In this way, substituting Eqs. \eqref{solu-Phi-0} and \eqref{Solu-Phin} in Eq. \eqref{Phi-Gen}, with $\varepsilon=1$, we find the solution for the potential related to the problem of a grounded conducting corrugated cylinder placed in a uniform electric field.
Note that, to find Eq. \eqref{Solu-Phin}, we did not make calculations in Fourier space, which makes this example even more simple than that of a point charge in the presence of a corrugated surface (discussed in Sec. \ref{review-clinton}),
becoming it a good pedagogical model to introduce this perturbative approach for corrugated surfaces in an electromagnetism course.

We can write explicitly the perturbative corrections $\Phi^{(n)}$ up to the third order:
\begin{widetext}
\begin{align}
	\Phi^{(1)}\left(\rho,\theta\right) & =\sum_{l=1}^{\infty}\frac{2E_{0}\rho}{\pi}\left(\frac{a}{\rho}\right)^{l+1}\int_{0}^{2\pi}d\bar{\theta}_{1}\cos\left[l\left(\bar{\theta}_{1}-\theta\right)\right]\cos\left(\bar{\theta}_{1}\right)\frac{h\left(\bar{\theta}_{1}\right)}{a},	\label{Phi1}\\
	\nonumber \\
	\Phi^{(2)}\left(\rho,\theta\right) & =\sum_{l=1}^{\infty}\frac{E_{0}\rho}{\pi}\left(\frac{a}{\rho}\right)^{l+1}\int_{0}^{2\pi}d\bar{\theta}_{2}\cos\left[l\left(\bar{\theta}_{2}-\theta\right)\right]\nonumber \\ \label{Phi2}
	& \times\left\{ \frac{2l}{\pi}\int_{0}^{2\pi}d\bar{\theta}_{1}\cos\left[l\left(\bar{\theta}_{1}-\bar{\theta}_{2}\right)\right]\cos\left(\bar{\theta}_{1}\right)\frac{h\left(\bar{\theta}_{2}\right)h\left(\bar{\theta}_{1}\right)}{a^{2}}-\cos\left(\bar{\theta}_{2}\right)\frac{h\left(\bar{\theta}_{2}\right)^{2}}{a^{2}}\right\} ,\\
	\nonumber \\
	\Phi^{(3)}\left(\rho,\theta\right) & =\sum_{l=1}^{\infty}\frac{E_{0}\rho}{\pi}\left(\frac{a}{\rho}\right)^{l+1}\int_{0}^{2\pi}d\bar{\theta}_{3}\cos\left[l\left(\bar{\theta}_{3}-\theta\right)\right]\left\{ \frac{l}{\pi}\int_{0}^{2\pi}d\bar{\theta}_{2}\cos\left[l\left(\bar{\theta}_{2}-\bar{\theta}_{3}\right)\right]\right.\nonumber \\
	& \times\left[\frac{2l}{\pi}\int_{0}^{2\pi}d\bar{\theta}_{1}\cos\left[l\left(\bar{\theta}_{1}-\bar{\theta}_{2}\right)\right]\cos\left(\bar{\theta}_{1}\right)\frac{h\left(\bar{\theta}_{3}\right)h\left(\bar{\theta}_{2}\right)h\left(\bar{\theta}_{1}\right)}{a^{3}}-\cos\left(\bar{\theta}_{2}\right)\frac{h\left(\bar{\theta}_{3}\right)h\left(\bar{\theta}_{2}\right)^{2}}{a^{3}}\right]\nonumber \\	\label{Phi3}
	& -\left.\frac{l\left(l+1\right)}{\pi}\int_{0}^{2\pi}d\bar{\theta}_{1}\cos\left[l\left(\bar{\theta}_{1}-\bar{\theta}_{3}\right)\right]\cos\left(\bar{\theta}_{1}\right)\frac{h\left(\bar{\theta}_{3}\right)^{2}h\left(\bar{\theta}_{1}\right)}{a^{3}}+\cos\left(\bar{\theta}_{3}\right)\frac{h\left(\bar{\theta}_{3}\right)^{3}}{a^{3}}\right\} .
\end{align}
\end{widetext}
In all these terms, one can identify the ratio $|h|/a$ as the perturbative parameter (remembering that we started by considering $\text{max}|h(\theta)|/a\ll 1$), which means that $\Phi^{(1)}\propto |h|/a$, $\Phi^{(2)}\propto (|h|/a)^2$, and $\Phi^{(3)}\propto (|h|/a)^3$.
Thus, as the ratio $|h|/a$ decreases, the faster our perturbative solution converges.
Besides this, note that, in each one of these perturbative corrections, 
the ratio $a/\rho$ controls the convergence of the series on $l$, which means that as the ratio $a/\rho$ decreases, the faster these series converge. 
Lastly, it is important to remark that the solution found here is general, in the sense that $h(\theta)$ represents a general function. 
In this way, in the next section we consider some applications of our results.

\section{Cylinder with a slight enhancement in its radius}
\label{radius sigthly enhanced}

We can verify the validity of our calculations by choosing a constant corrugation profile, $h(\theta)=\delta$, which should yields the solution for a cylinder of radius $a+\delta$.
From Eq. \eqref{Solu-Phin}, we obtain, for this case, that:
\begin{align}
\label{delta1}
\Phi^{(1)}\left(\rho,\theta\right) & =2\delta E_{0}\left(\frac{a}{\rho}\right)\cos\left(\theta\right),
\\
\label{delta2}
\Phi^{(2)}\left(\rho,\theta\right) & =\frac{\delta^{2}}{\rho}E_{0}\cos\left(\theta\right),
\\
\label{delta3}
\Phi^{(i)}\left(\rho,\theta\right) & =0 \quad (i \geq 3).
\end{align}
Therefore, Eq. \eqref{Phi-Gen} (with $\varepsilon=1$) can be written as
\begin{eqnarray}
\label{teste_veracidade}
\Phi\left(\rho,\theta\right)=-E_{0}\rho\cos\left(\theta\right)\left(1-\frac{\left(a+\delta\right)^{2}}{\rho^{2}}\right),
\end{eqnarray} 
which, as expected, corresponds to the result for a cylinder of radius $a+\delta$.

\section{Sinusoidal corrugated cylinder}
\label{application Cylinder corrgated}

As another application, let us consider a sinusoidal corrugated cylinder, whose profile is described by
\begin{equation}
\label{h}
h(\theta)=\frac{\delta}{2}[1+\cos(\nu\theta)],
\end{equation}
where $\nu \in \mathbb{N}$.
Note that, when $\nu=0$, Eq. \eqref{h} describes the case of a cylinder with a slight enhancement in its radius.

\begin{figure}[h]
\centering
\subfigure[\label{Plot-Pot-liso}]{\epsfig{file=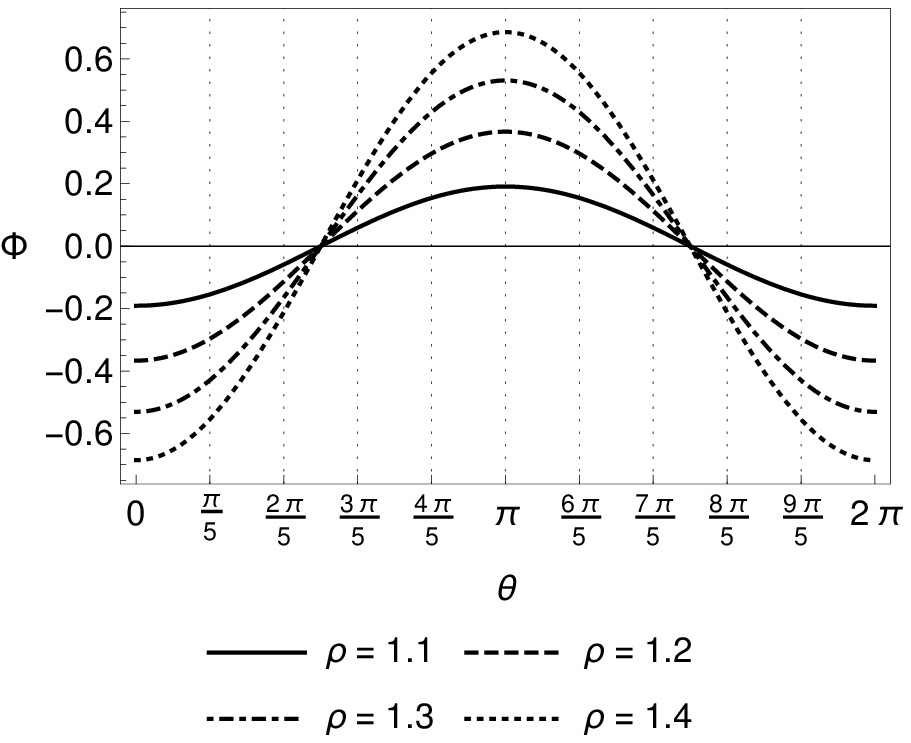,  width=0.8 \linewidth}}
\subfigure[\label{Plot-Pot-corrugado}]{\epsfig{file=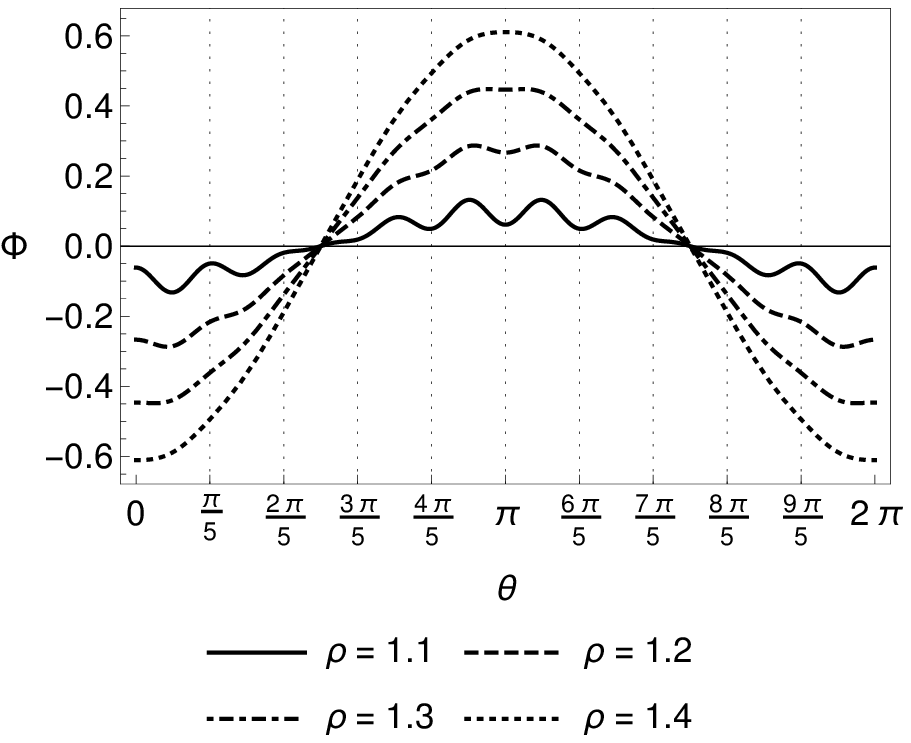,  width=0.8 \linewidth}}
\caption{{ 
Behavior of the electrostatic potential, as a function of $\theta$, for the configurations of (a) a non-corrugated cylinder, and (b) a sinusoidal corrugated one.
We considered $a=1$, $\delta=0.1$, $\nu=10$, $E_{0}=1$.
In (b), each tick on the horizontal axis represent a corrugation peak.
}}
\label{Plot-Pot}
\end{figure}

In this case, we calculate an approximate solution for $\Phi$ up to first-order in $h$, which is given by
\begin{equation}
\Phi(\rho,\theta)\approx\Phi^{(0)}(\rho,\theta)+\Phi^{(1)}(\rho,\theta),
\end{equation}
with $\Phi^{(0)}$ given by Eq. \eqref{solu-Phi-0}.
To calculate $\Phi^{(1)}$, we use Eq. \eqref{h} in Eq. \eqref{Phi1} and obtain
\begin{align}
	\nonumber
\Phi^{(1)}\left(\rho,\theta\right) & =\frac{E_{0}\delta}{\pi}\sum_{l=1}^{\infty}\left(\frac{a}{\rho}\right)^{l}\\
 & \times\int_{0}^{2\pi}d\bar{\theta}\cos\left[l\left(\bar{\theta}-\theta\right)\right]\left[1+\cos\left(\nu\bar{\theta}\right)\right]\cos\left(\bar{\theta}\right).
\end{align}
The above integral is given by
\begin{align}
\nonumber
\int_{0}^{2\pi}d\bar{\theta}\cos\left[l\left(\bar{\theta}-\theta\right)\right]\left[1+\cos\left(\nu\bar{\theta}\right)\right]\cos\left(\bar{\theta}\right)\\
=\pi\cos\left(l\theta\right)\delta_{l1}+\frac{\pi}{2}\cos\left(l\theta\right)(\delta_{l,|1-\nu|} & +\delta_{l,\nu+1}).
\end{align}
From this, we find that $\Phi^{(1)}$ is given by
\begin{align}
\Phi^{(1)}(\rho,\theta) & =\frac{\delta E_{0}a}{\rho}\cos(\theta)+\frac{\delta E_{0}}{2}\left\{ \left(\frac{a}{\rho}\right)^{\nu+1}\cos[(\nu+1)\theta]\right.\nonumber \\ \label{qualitative}
 & \left.+\left(\frac{a}{\rho}\right)^{|\nu-1|}\cos[|\nu-1|\theta]-\delta_{\nu1}\right\} 
\end{align}
One can note from this equation that, for $\nu=0$, we recover Eq. \eqref{delta1}, as expected.
In Fig. \ref{Plot-Pot}, we show the behavior of the potential for the case of a non-corrugated cylinder and compare it with the case of a corrugated one.
In this figure, one can note that the magnitude of the potential for a corrugated cylinder oscillates close to that calculated for a non-corrugated one. 
As $\rho$ increases, we can see that the corrugation effects vanish and the potential becomes similar to the one for the non-corrugated case.

\begin{figure}[h]
\centering
\subfigure[\label{Plot-Stream-liso}]{\epsfig{file=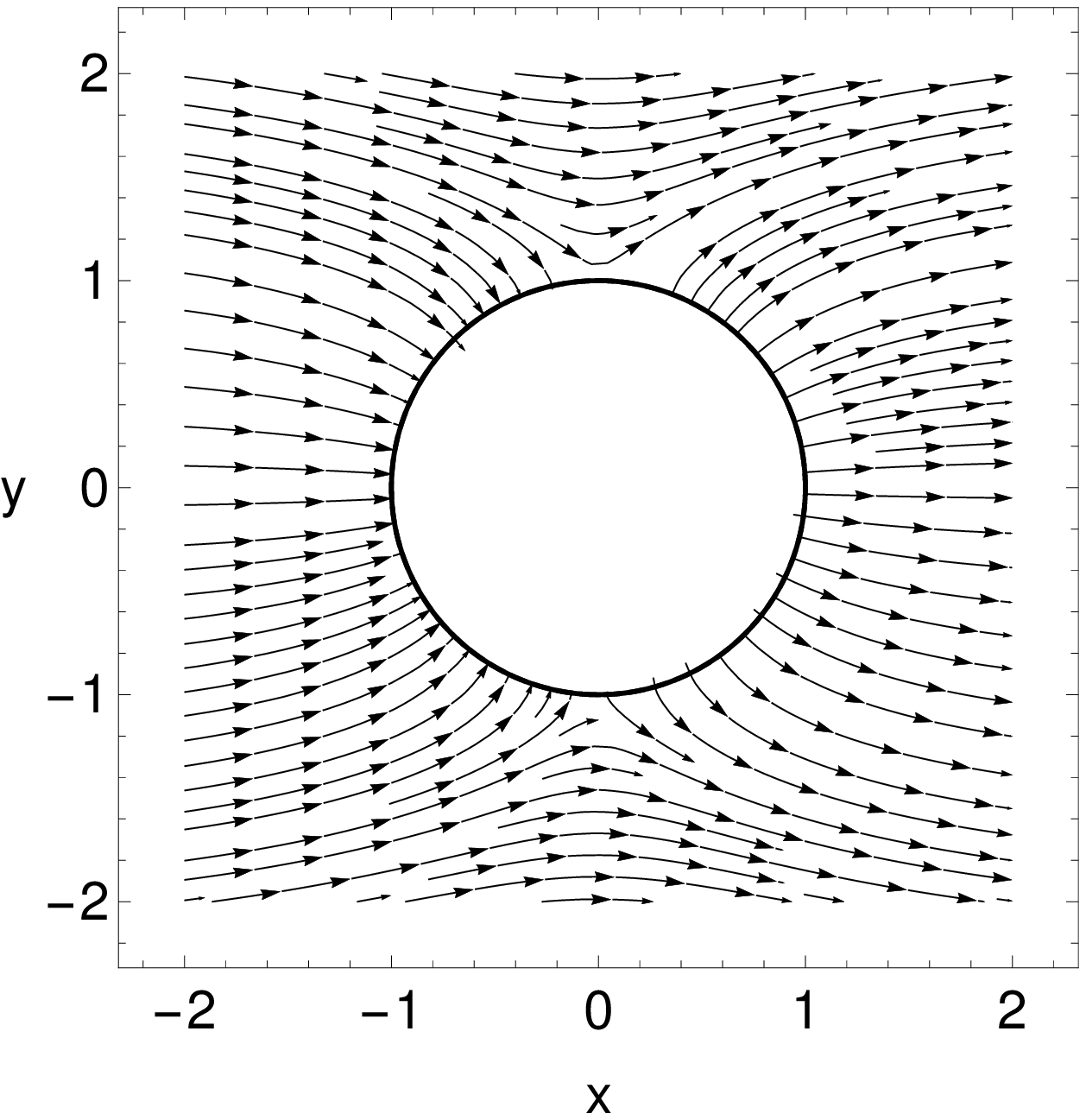,  width=0.7 \linewidth}}
\subfigure[\label{Plot-Stream-corrugado}]{\epsfig{file=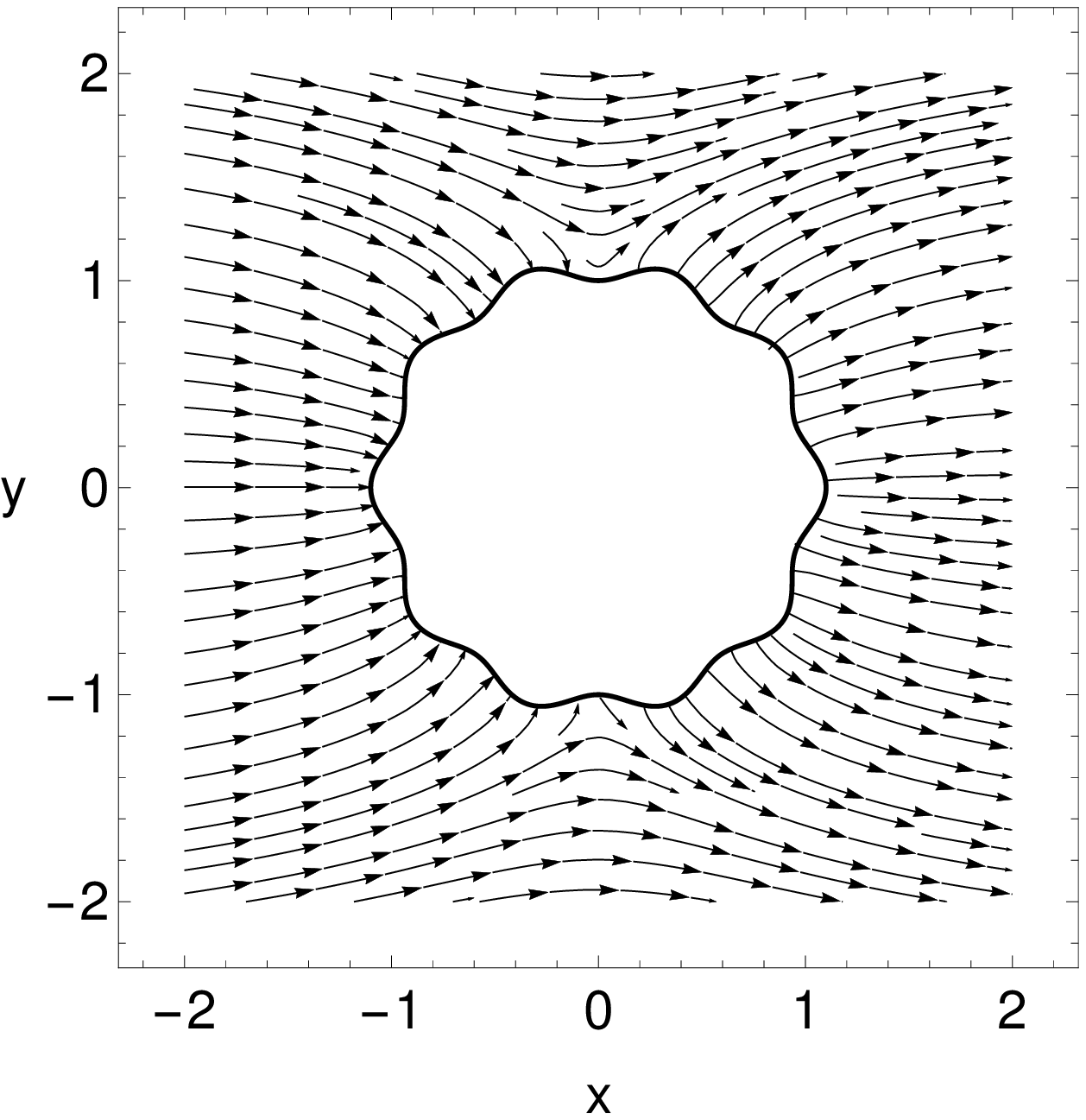,  width=0.7 \linewidth}}
\caption{{
Behavior of the electric field lines, along the plane $z=0$, for the configurations of (a) a non-corrugated cylinder, and (b) a sinusoidal corrugated one.
We considered $a=1$, $\delta=0.1$, $\nu=10$, $E_{0}=1$.
}}
\label{Plot-Stream}
\end{figure}

\begin{figure}[h]
\centering
\subfigure[\label{Plot-Charge-10}]{\epsfig{file=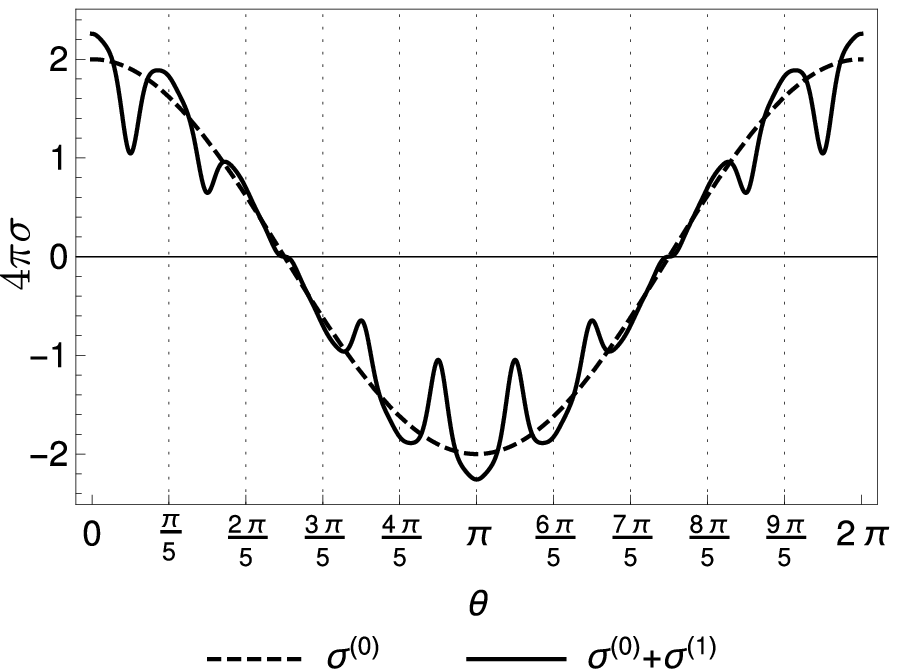,  width=0.8 \linewidth}}
\subfigure[\label{Plot-Charge-11}]{\epsfig{file=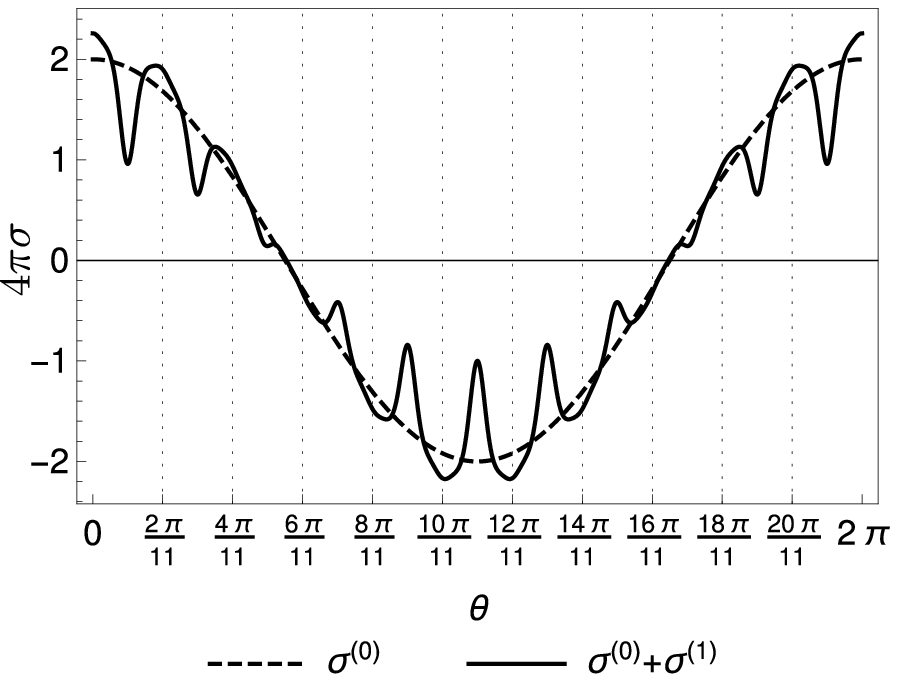,  width=0.8 \linewidth}}
\caption{{
Behavior of the induced surface charge density $\sigma$, along the plane $z=0$, on a cylinder placed in a uniform electrostatic field $\textbf{E}_{0}= \hat{{\bf x}}$.
In both figures, $\sigma^{(0)}$ (dashed line) represents the case of a non-corrugated cylinder, whereas $\sigma^{(0)}+\sigma^{(1)}$ (solid line) represents the case of a sinusoidal corrugated cylinder calculated up to first perturbative order.
We consider $a=1$ and $\delta=0.1$, with $\nu=10$ in (a), and $\nu=11$ in (b).
Each tick on the horizontal axis represent a corrugation peak of the corrugated cylinder.
}}
\label{Plot-Charge}
\end{figure}

From the obtained approximate solution for the electrostatic potential, we can also calculate the electric field $\textbf{E}$ for this configuration, and the induced surface charge density on the cylinder [Eq. \eqref{densidade-carga}].
In this way, in Figs. \ref{Plot-Stream} and \ref{Plot-Charge}, we show, respectively, the behavior of the electric field lines and the induced surface charge density for this case, and compare them with those related to a non-corrugated cylinder of radius $a$.
In Fig. \ref{Plot-Stream}, we can see that, for the considered values of $a$, $\nu$ and $E_0$, the electric field lines are practically perpendicular to the surface, which shows that, for this case, we have an acceptable approximate solution.
Lastly, in Fig. \ref{Plot-Charge}, one can note that, due to the oscillatory profile of the sinusoidal surface, the induced surface charge density for this case oscillates around that of a non-corrugated cylinder.
%
%
%

		
\section{Final Remarks}
\label{final-remarks}
In the present paper, we discussed the introduction, via perturbative approach, 
of corrugated surfaces in basic electromagnetism problems.
We started, in Sec. \ref{review-point-plane}, discussing a very common exercise applied in introductory courses, 
namely to solve the Poisson's equation for a point charge in the presence of an infinity
perfectly conducting planar surface, whose solution is usually obtained by image method.
The introduction of corrugations in this model was discussed by means
of a detailed pedagogical review of the calculations of Clinton, Esrick and Sacks \cite{Clinton-PRBII-1985}.
As an application of these calculations, we investigated the case of a sinusoidal corrugated surface, and showed how the presence of corrugations affect the behavior of the electric field lines and the induced surface charge density (see Figs. \ref{fig:linhas-campo} and \ref{fig:densidade-carga}).
Moreover, in Sec. \ref{corrugated-cylinder}, we extended this perturbative approach and introduced corrugation in another common problem in electromagnetism courses, namely the neutral infinity grounded conducting cylinder placed in a uniform electric field.
As an application of our calculations, we investigated the case of a sinusoidal corrugated cylinder, and showed how the presence of corrugations affect the behavior of the electric field lines and the induced surface charge density (see Figs. \ref{Plot-Stream} and \ref{Plot-Charge}).
We showed that the problem of a corrugated cylinder does not require calculations in Fourier space, which makes it even more simple than that of a point charge in the presence of a corrugated surface, becoming a good pedagogical model to introduce this perturbative approach for corrugated surfaces in an electromagnetism course.
%

\appendix
\section{Some calculations used in Sec. I}

\subsection{Transformation of $\left.G^{(n)}\left(\textbf{r}_{\parallel},\textbf{r}_{\parallel}^{\prime};z,z^{\prime}\right)\right|_{z=0}$ to Fourier Space} \label{transf-of-gn}

From the boundary condition
\begin{align}
\nonumber
&\left.G^{(n)}\left(\textbf{r}_{\parallel},\textbf{r}_{\parallel}^{\prime};z,z^{\prime}\right)\right|_{z=0}=
\\
&-\sum_{m=1}^{n}\frac{\left[h(\textbf{r}_\parallel)\right]^{m}}{m!}\left.\frac{\partial^{m}}{\partial z^{m}}G^{(n-m)}\left(\textbf{r}_{\parallel},\textbf{r}_{\parallel}^{\prime};z,z^{\prime}\right)\right|_{z=0},
\label{Gn-Boundary-apendice}
\end{align}
if we write $G^{(n-m)}\left(\textbf{r}_{\parallel}, \textbf{r}_{\parallel}^{\prime}; z,z^{\prime}\right)$ as an inverse Fourier transformation, one can obtain
\begin{align}
	\nonumber
	&\left.G^{(n)}\left(\textbf{r}_{\parallel},\textbf{r}_{\parallel}^{\prime};z,z^{\prime}\right)\right|_{z=0}=-\sum_{m=1}^{n}\frac{\left[h(\textbf{r}_\parallel)\right]^{m}}{m!}
	\\ \label{this one}
	&\times\int\frac{d^{2}\textbf{k}^{\prime}}{\left(2\pi\right)^{2}} e^{i\textbf{k}^{\prime}\cdot\textbf{r}_{\parallel}}\left.\frac{\partial^{m}}{\partial z^{m}}\tilde{G}^{(n-m)}\left(\textbf{k}^{\prime},\textbf{r}_{\parallel}^{\prime};z^{\prime},z\right)\right|_{z=0}.
\end{align}
By applying a Fourier transformation in Eq. \eqref{this one}, we obtain
\begin{align} \left.G^{(n)}\left(\textbf{k},\textbf{r}_{\parallel}^{\prime};z,z^{\prime}\right)\right|_{z=0} &=-\sum_{m=1}^{n}\frac{1}{m!}\int\frac{d^{2}\textbf{k}^{\prime}}{\left(2\pi\right)^{2}} \nonumber \\
 & \times\left(\int d^{2}\textbf{r}_{\parallel}\left[h(\textbf{r}_{\parallel})\right]^{m}e^{-i\left(\textbf{k}-\textbf{k}^{\prime}\right)\cdot\textbf{r}_{\parallel}}\right) \nonumber  \\
 & \times\left.\frac{\partial^{m}}{\partial z^{m}}\tilde{G}^{(n-m)}\left(\textbf{k}^{\prime},\textbf{r}_{\parallel}^{\prime};z^{\prime},z\right)\right|_{z=0}].
\end{align}
Note that the term inside the parenthesis can be written as
\begin{equation}
\tilde{h}_{m}(\textbf{k}-\textbf{k}^{\prime})=\int d^{2}\textbf{r}_{\parallel}e^{-i(\textbf{k}-\textbf{k}^{\prime})\cdot \textbf{r}_{\parallel}}\left[h(\textbf{r}_\parallel)\right]^{m}.
\end{equation}
In this way, the boundary condition given by Eq. \eqref{Gn-Boundary-apendice} can be written in the Fourier space as
\begin{align} 
\nonumber
\left.\tilde{G}^{(n)}\left(\textbf{k},\textbf{r}_{\parallel}^{\prime};z,z^{\prime}\right)\right|_{z=0} &= -\sum_{m=1}^{n}\frac{1}{m!}\int \frac{d^{2}\textbf{k}^{\prime}}{\left(2\pi\right)^{2}}\tilde{h}_{m}\left(\textbf{k}-\textbf{k}^{\prime}\right)
\\
&\times  \left.\frac{\partial^{m}}{\partial z^{m}}\tilde{G}^{(n-m)}\left(\textbf{k}^{\prime},\textbf{r}_{\parallel}^{\prime};z^{\prime},z\right)\right|_{z=0}.
\end{align}
%

\subsection{Solution of $\tilde{G}^{(0)}\left(\textbf{k},\textbf{r}_{\parallel}^{\prime};z,z^{\prime}\right)$}
\label{calc-of-g0}

The equation to be solved is
\begin{equation}\label{g0}
	\left(\frac{\partial^{2}}{\partial z^{2}}-\left|\textbf{k}\right|^{2}\right)\tilde{G}^{(0)}\left(\textbf{k},\textbf{r}_{\parallel}^{\prime};z,z^{\prime}\right)=-4\pi e^{-i\textbf{k}\cdot\textbf{\textbf{r}}_{\parallel}^{\prime}}\delta\left(z-z^{\prime}\right),
\end{equation}
under the boundary conditions
\begin{align}
	\label{bc-1surf}
	\left.\tilde{G}^{(0)}\left(\textbf{k},\textbf{r}_{\parallel}^{\prime};z,z^{\prime}\right)\right|_{z=0}  &=0,
	\\
	\label{bc-2inf}
	\left. \tilde{G}^{(0)}\left(\textbf{k},\textbf{r}_{\parallel}^{\prime};z,z^{\prime}\right)\right|_{z\to \infty} &=0.
\end{align}
The solution of Eq. \eqref{g0} is described in the regions for $z<z^{\prime}$ and $z>z^{\prime}$, so that one can obtain
\begin{align}
	\tilde{G}^{(0)}_{<}\left(\textbf{k},\textbf{r}_{\parallel}^{\prime};z,z^{\prime}\right) & =A_{<}e^{\vert\textbf{k}\vert z}+B_{<}e^{-\vert\textbf{k}\vert z},\\
	\tilde{G}^{(0)}_{>}\left(\textbf{k},\textbf{r}_{\parallel}^{\prime};z,z^{\prime}\right) & =A_{>}e^{\vert\textbf{k}\vert z}+B_{>}e^{-\vert\textbf{k}\vert z}.
\end{align}
From the boundary conditions, Eqs. \eqref{bc-2inf} and \eqref{bc-1surf}, one obtains
\begin{align}\label{g0<}
	\tilde{G}^{(0)}_{<}\left(\textbf{k},\textbf{r}_{\parallel}^{\prime};z,z^{\prime}\right) & =A_{<}\left(e^{\vert\textbf{k}\vert z}-e^{-\vert\textbf{k}\vert z}\right),\\ \label{g0>}
	\tilde{G}^{(0)}_{>}\left(\textbf{k},\textbf{r}_{\parallel}^{\prime};z,z^{\prime}\right) & =B_{>}e^{-\vert\textbf{k}\vert z}.
\end{align}
Furthermore, by knowing that the Green's function is continuous in $z = z^{\prime}$, we can write 
\begin{equation}
\tilde{G}^{(0)}_{<}\left(\textbf{k},z=z^{\prime};\textbf{r}_{\parallel}^{\prime},z^{\prime}\right)=\tilde{G}^{(0)}_{>}\left(\textbf{k},z=z^{\prime};\textbf{r}_{\parallel}^{\prime},z^{\prime}\right).
\end{equation}
From this we find
\begin{equation}\label{B>1}
	B_{>}=A_{<}e^{\vert\textbf{k}\vert z^{\prime}}\left(e^{\left|\textbf{k}\right|z^{\prime}}-e^{-\vert\textbf{k}\vert z^{\prime}}\right).
\end{equation}
We can also obtain the discontinuity of the Green's function by integrating Eq. \eqref{g0} around $z=z^{\prime}$, so that
\begin{align}
	\nonumber
	&\int_{z^{\prime}-\epsilon}^{z^{\prime}+\epsilon}\left(\frac{d^{2}}{dz^{2}}-\vert\textbf{k}\vert^{2}\right)\tilde{G}^{(0)}\left(\textbf{k},\textbf{r}_{\parallel}^{\prime};z,z^{\prime}\right)dz=
	\\
	&-4\pi e^{-i\textbf{k}\cdot\textbf{r}_{\parallel}^{\prime}} \int_{z^{\prime}-\epsilon}^{z^{\prime}+\epsilon} \delta\left(z-z^{\prime}\right)dz,
\end{align}
or yet
\begin{align}
	\nonumber
	&\left(\frac{d\tilde{G}^{(0)}}{dz}\right)_{z^{\prime}+\epsilon}-\left(\frac{d\tilde{G}^{(0)}}{dz}\right)_{z^{\prime}-\epsilon}
	\\
	&-\vert\textbf{k}\vert^{2}\int_{z^{\prime}-\epsilon}^{z^{\prime}+\epsilon}\tilde{G}^{(0)}\left(\textbf{k},\textbf{r}_{\parallel}^{\prime};z,z^{\prime}\right)dz=-4\pi e^{-i\textbf{k}\cdot\textbf{r}_{\parallel}^{\prime}}.
\end{align}
Since the Green's function is continuous at $z=z^{\prime}$, in the limit $\epsilon\to0$, one obtains
\begin{eqnarray}
	\left(\frac{d\tilde{G}^{(0)}_{>}}{dz}\right)_{z=z^{\prime}}-\left(\frac{d\tilde{G}^{(0)}_{<}}{dz}\right)_{z=z^{\prime}}=-4\pi e^{-i\textbf{k}\cdot\textbf{r}_{\parallel}^{\prime}}.
\end{eqnarray}
Substituting Eqs. \eqref{g0<} and \eqref{g0>} in this equation, we obtain
\begin{equation}\label{B>2}
	B_{>}e^{-\vert\textbf{k}\vert z^{\prime}}+A_{<}\left(e^{\vert\textbf{k}\vert z^{\prime}}+e^{-\vert\textbf{k}\vert z^{\prime}}\right)=\frac{4\pi}{\vert\textbf{k}\vert}e^{-i\textbf{k}\cdot\textbf{r}_{\parallel}^{\prime}}.
\end{equation}
From Eqs. \eqref{B>1} and \eqref{B>2}, one obtains that $A_{<}$ and $B_{>}$ are given by: 
\begin{align}
	A_{<} & =\frac{2\pi}{\vert\textbf{k}\vert}e^{-\vert\textbf{k}\vert z^{\prime}}e^{-i\textbf{k}\cdot\textbf{r}_{\parallel}^{\prime}},\\
	B_{>} & =\frac{2\pi}{\vert\textbf{k}\vert}\left(e^{\vert\textbf{k}\vert z^{\prime}}-e^{-\vert\textbf{k}\vert z^{\prime}}\right)e^{-i\textbf{k}\cdot\textbf{r}_{\parallel}^{\prime}}.
\end{align}
Therefore, Eqs. \eqref{g0<} and \eqref{g0>} become:
\begin{align}
	\tilde{G}^{(0)}_{<}\left(\textbf{k},\textbf{r}_{\parallel}^{\prime};z,z^{\prime}\right) & =\frac{2\pi}{\vert\textbf{k}\vert}e^{-i\textbf{k}\cdot\textbf{r}_{\parallel}^{\prime}}\left[e^{-\vert\textbf{k}\vert\left(z^{\prime}-z\right)}-e^{-\vert\textbf{k}\vert\left(z+z^{\prime}\right)}\right],\\
	\tilde{G}^{(0)}_{>}\left(\textbf{k},\textbf{r}_{\parallel}^{\prime};z,z^{\prime}\right) & =\frac{2\pi}{\vert\textbf{k}\vert}e^{-i\textbf{k}\cdot\textbf{r}_{\parallel}^{\prime}}\left[e^{-\vert\textbf{k}\vert\left(z-z^{\prime}\right)}-e^{-\vert\textbf{k}\vert\left(z+z^{\prime}\right)}\right].
\end{align}
These two equations can be taken together, so that the solution of $\tilde{G}^{(0)}\left(\textbf{k},\textbf{r}_{\parallel}^{\prime};z,z^{\prime}\right)$ can be written as 
\begin{equation}
\tilde{G}^{(0)}\left(\textbf{k},\textbf{r}_{\parallel}^{\prime};z,z^{\prime}\right)=\frac{2\pi}{\left|\textbf{k}\right|}e^{-i\textbf{k}\cdot\textbf{r}_{\parallel}^{\prime}} \left[e^{-\left|\textbf{k}\right|\left|z-z^{\prime}\right|}-e^{-\left|\textbf{k}\right|\left(z+z^{\prime}\right)}\right].
\end{equation}
%

\subsection{Solution of $\tilde{G}^{(n)}\left(\textbf{k},\textbf{r}_{\parallel}^{\prime};z,z^{\prime}\right)$}\label{calc-of-gn}
The equation to be solved is
\begin{equation}\label{Helmholtz}
\left(\frac{\partial^{2}}{\partial z^{2}}-\left|\textbf{k}\right|^{2}\right)\tilde{G}^{(n)}\left(\textbf{k},\textbf{r}_{\parallel}^{\prime};z^{\prime},z\right)=0\qquad (n\geq1),
\end{equation}
under the boundary conditions given by
\begin{align}
\tilde{G}^{(n)}(\textbf{k},\textbf{r}_{\parallel}^{\prime};z,z^{\prime})\vert_{z=0} &=  -\sum_{m=1}^{n}\int\frac{d^{2}\textbf{k}^{\prime}}{(2\pi)^{2}}\frac{\tilde{h}_{m}(\textbf{k}-\textbf{k}^{\prime})}{m!}
\nonumber \\ 
\label{Gn-Boundary-Fourier-apendice}
&\times\frac{\partial^{m}}{\partial z^{m}}\tilde{G}^{(n-m)}(\textbf{k}^{\prime},\textbf{r}_{\parallel}^{\prime};z,z^{\prime})\vert_{z=0}, \\
\label{inf-bc}
\left.\tilde{G}^{(n)}\left(\textbf{k},\textbf{r}_{\parallel}^{\prime};z^{\prime},z\right)\right|_{z\to \infty}  &= 0.
\end{align} 
The solution of Eq. \eqref{Helmholtz} is given by
\begin{equation}\label{equation-the}
	\tilde{G}^{(n)}\left(\textbf{k},\textbf{r}_{\parallel}^{\prime};z^{\prime},z\right)=Ae^{-\left|\textbf{k}\right|z}+Be^{\left|\textbf{k}\right|z}.
\end{equation}
From Eq. \eqref{inf-bc}, one obtains
\begin{equation}
	\tilde{G}^{(n)}\left(\textbf{k},\textbf{r}_{\parallel}^{\prime};z^{\prime},z\right)=Ae^{-\left|\textbf{k}\right|z}\label{eq:A6}.
\end{equation}
Besides this, from Eq. \eqref{Gn-Boundary-Fourier-apendice}, one obtains
\begin{align}
A & =-\sum_{m=1}^{n}\int\frac{d^{2}\textbf{k}^{\prime}}{(2\pi)^{2}}\frac{\tilde{h}_{m}(\textbf{k}-\textbf{k}^{\prime})}{m!}\nonumber\\
 & \times\frac{\partial^{m}}{\partial z^{m}}\tilde{G}^{(n-m)}(\textbf{k}^{\prime},\textbf{r}_{\parallel}^{\prime};z,z^{\prime})\vert_{z=0}.
\end{align}
Therefore, the solution of $\tilde{G}^{(n)} \left( \textbf{k} , \textbf{r}_{\parallel}^{\prime} ; z, z^{\prime} \right)$ can be written as 
\begin{align}
	\nonumber
	\tilde{G}^{(n)}\left(\textbf{k},\textbf{r}_{\parallel}^{\prime};z^{\prime},z\right)&=
	-e^{-\left|\textbf{k}\right|z}\sum_{m=1}^{n}\int\frac{d^{2}\textbf{k}^{\prime}}{\left(2\pi\right)^{2}}\frac{\tilde{h}_{m}\left(\textbf{k}-\textbf{k}^{\prime}\right)}{m!}
	\\
	&\times\left.\frac{\partial^{m}}{\partial z^{m}}\tilde{G}^{(n-m)}\left(\textbf{k}^{\prime},\textbf{r}_{\parallel}^{\prime};z^{\prime},z\right)\right|_{z=0}.
\end{align}
%

\begin{acknowledgments}
	The authors thank Alessandra Braga, Marcelo Lima, Stanley Coelho, and Van Sérgio Alves, for a careful reading of this paper and fruitful discussions.
	A.P.C. was supported by the Conselho Nacional de Desenvolvimento Cient\'{i}fico e Tecnol\'{o}gico - Brasil (CNPq) - Brasil.
	A.P.C., L.Q. and E.C.M.N. were supported by the Coordena\c{c}\~{a}o de Aperfei\c{c}oamento de Pessoal de N\'{i}vel Superior (CAPES) - Brasil, Finance Code 001.
\end{acknowledgments}
		

%
		
\end{document}